\documentclass[12pt]{iopart}
\expandafter\let\csname equation*\endcsname\relax
\expandafter\let\csname endequation*\endcsname\relax
\usepackage{amsfonts} 
\usepackage{amsmath}
\usepackage{iopams}
\usepackage{amssymb}
\usepackage{mathtools}
\usepackage{cancel} 
\usepackage{bm}
\usepackage[usenames, dvipsnames]{color} 
\usepackage{dcolumn}
\usepackage{epic} 
\usepackage{epsfig}
\usepackage{feynmp}
\usepackage{graphicx}
\usepackage{grffile}
\usepackage[breaklinks,colorlinks = true,linkcolor = red,urlcolor=blue,citecolor=red]{hyperref}
\usepackage{mathrsfs}
\usepackage{soul}
\usepackage{wrapfig}
\usepackage{xy} 
\usepackage{xcolor}
\usepackage{amsmath,amssymb,amsfonts,amsthm}
\usepackage[ascii]{inputenc}
\usepackage{makecell}
\allowdisplaybreaks

\newcommand{\be}{\begin{equation}}
	\newcommand{\ee}{\end{equation}}
\newcommand{\ba}{\begin{aligned}}
	\newcommand{\ea}{\end{aligned}}
\newcommand{\bea}{\begin{eqnarray}}
	\newcommand{\eea}{\end{eqnarray}}

\newcommand{\bx}{{\bf x}}

\newcommand{\red}{\textcolor{red}}

\graphicspath{{images/}{../Figures}}

\begin{document}

\title{Linear statistics at the microscopic scale for the 2D Coulomb gas}


\author{Pierre Le Doussal}
\address{Laboratoire de Physique de l'Ecole Normale Sup\'erieure, CNRS, ENS and PSL Universit\'e, Sorbonne Universit\'e, Universit\'e Paris Cit\'e,
24 rue Lhomond, 75005 Paris, France}
\author{Gr\'egory Schehr}
\address{Sorbonne Universit\'e, Laboratoire de Physique Th\'eorique et Hautes Energies, CNRS UMR 7589, 4 Place Jussieu, 75252 Paris Cedex 05, France}



\begin{abstract} 
We consider the classical Coulomb gas in two dimensions at the inverse temperature $\beta=2$,
confined within a droplet of radius $R$ by a rotationally invariant potential $U(r)$. For $U(r)\sim r^2$ this describes the 
eigenvalues of the complex Ginibre ensemble of random matrices. We study linear statistics of the form 
${\cal L}_N = \sum_{i=1}^N f(|\bx_i|)$, 
where $\bx_i$'s
are the positions of the $N$ particles, in the large $N$ limit 
with $R=O(1)$.
It is known that for smooth functions $f(r)$ the
variance ${\rm  Var} \,{\cal L}_N= O(1)$, while for the indicator function $f(r)= {\mathbb I}_{0<r<\hat r}$ with $0<\hat r \leq R$, relevant 
for the disk counting statistics,
all cumulants of ${\cal L}_N$ of order $q \geq 2$ behave as $\sim \sqrt{N}$. In addition, for smooth functions, it was shown recently that the cumulants of ${\cal L}_N$ 
of order $q \geq 3$ scale as $O(N^{2-q})$. Surprisingly it was found that they
depend only on $f'(|\bf x|)$ and its derivatives {\it evaluated exactly at the boundary of the droplet}. To understand this property, and interpolate between the two behaviors (smooth versus step-like), we study
the microscopic linear statistics given by $f(r) \to f_N(r) = \phi((r-\hat r) \sqrt{N}/\xi)$, which probes the fluctuations at the scale of the inter-particle distance. We compute the cumulants of ${\cal L}_N$ at large $N$ for a fixed shape function $\phi(u)$ at arbitrary $\xi$. 
For large $\xi$ they match the predictions for smooth functions 
which shows that the leading contribution in that case comes from a boundary
layer of size $1/\sqrt{N}$ near the boundary of the droplet. 
 Finally we show that the full probability
distribution of~${\cal L}_N$ take two distinct large deviation forms, in the regime ${\cal L}_N = O(\sqrt{N})$ and ${\cal L}_N =O(N)$ respectively. The transition between these two regimes is accompanied by the formation of a macroscopic hole in the distribution of charges. 
 We also discuss applications of our results to rotating fermions in a harmonic trap and to the Ginibre symplectic ensemble. 
\end{abstract}
\date{\today}
\maketitle

\maketitle

\section{Introduction}

The Coulomb gas in two-dimensions is a central model in statistical physics, with applications
ranging from Wigner crystals~\cite{Wigner34,Li21,Nelson2002} and melting of two dimensional solids \cite{Nelson2002} to 
topics in quantum Hall physics \cite{Cooper,Charles,oblak}, 
fermions in a rotating trap
\cite{Lacroix_rotating,Smith_rotating,Manas_rotating1,Manas_rotating2}, 
and, in
mathematics, around non-Hermitian random matrices \cite{Mehtabook,Forrester,Charlier22,lewin,Charlier23,Serfaty24,Byun25}. In particular the so-called 
``one component plasma'' involves $N$ particles interacting via the repulsive logarithmic
pairwise Coulomb potential in the presence of an external confining potential. The model
is studied at thermal equilibrium at inverse temperature $\beta$, and for simplicity
it is assumed that the external potential is rotationally invariant around the origin. 
In the special case $\beta=2$ 
the positions of the particles are in one to one correspondence with the eigenvalues
in the complex plane of complex normal random matrices \cite{Mehtabook,Forrester}. A well known example
is the complex Ginibre ensemble where the confining potential is harmonic. 
For such Coulomb gases the mean density of particles converges at large $N$ to a smooth asymptotic density profile with a finite support. This is usually called the {\it droplet}, which in the simplest (rotationally invariant) case
is a disk of radius~$R$. 

Because of the long range interactions, the fluctuations of the gas are non-trivial and they
have been much studied in statistical 
physics and in mathematics, see e.g. for reviews \cite{lewin,Dauxois}. 
An important class of observables in this context is called {\it linear statistics}, denoted by ${\cal L}_N$ and defined as ${\cal L}_N = \sum_{i=1}^N f({\bx_i})$, where the $\bx_i$'s denote the positions of the particles, 
and where $f(\bx)$ is an arbitrary function. Focusing here on rotationally invariant 
functions $f(\bx)=f(r)$, with $r= |\bx|$, an important example is
the indicator function $f(r)= {\mathbb I}_{0<r<\hat r}$, in which case 
${\cal L}_N$ is the number of particles inside the disk of radius $\hat r$ centered on the origin.
This is generically called the {\it full counting statistics} (FCS), which has been widely studied recently 
\cite{allez,Grabsch1,Xu,castillo,akemann1,akemann}.

The main challenge is to describe the universal behavior of the fluctuations in the large $N$ limit.
One usually scales the parameters such that the droplet radius remains $R=O(1)$ in the limit $N \to \infty$. With this scaling, 
it has been shown that for smooth functions $f(r)$ the
variance remains of order unity, i.e. ${\rm  Var} \,{\cal L}_N= O(1)$ \cite{RiderVirag2007,Ameur2011,Leble2018,BBNY,Flack2023}. 
In addition, the cumulants of ${\cal L}_N$ 
of order $q \geq 3$ were recently computed in \cite{BLMS23} and they were found to scale as $O(N^{2-q})$. Surprisingly it was also found \cite{BLMS23} that these higher order cumulants depend only on $f'(r)$ and its derivatives {\it evaluated exactly at the boundary of the droplet}.
On the other hand, it was also shown that for the FCS, all even cumulants of ${\cal L}_N$ of order $q \geq 2$ 
behave as $\sim \sqrt{N}$ for large $N$ \cite{Charles,Lacroix_rotating,Lambert2022}. 

It would therefore be very interesting to quantitatively understand the origin of the different $N$-dependence of these cumulants in the two cases. 
While it is clear that this difference stems from the contrasting behavior of the fluctuations at macroscopic
length scales (i.e. $O(1)$) and microscopic length scales (of order the interparticle distance
$1/\sqrt{N}$), the crossover between these two scales has not yet been analyzed in details.

\begin{figure}[t]
\centering
\includegraphics[width = 0.45\linewidth]{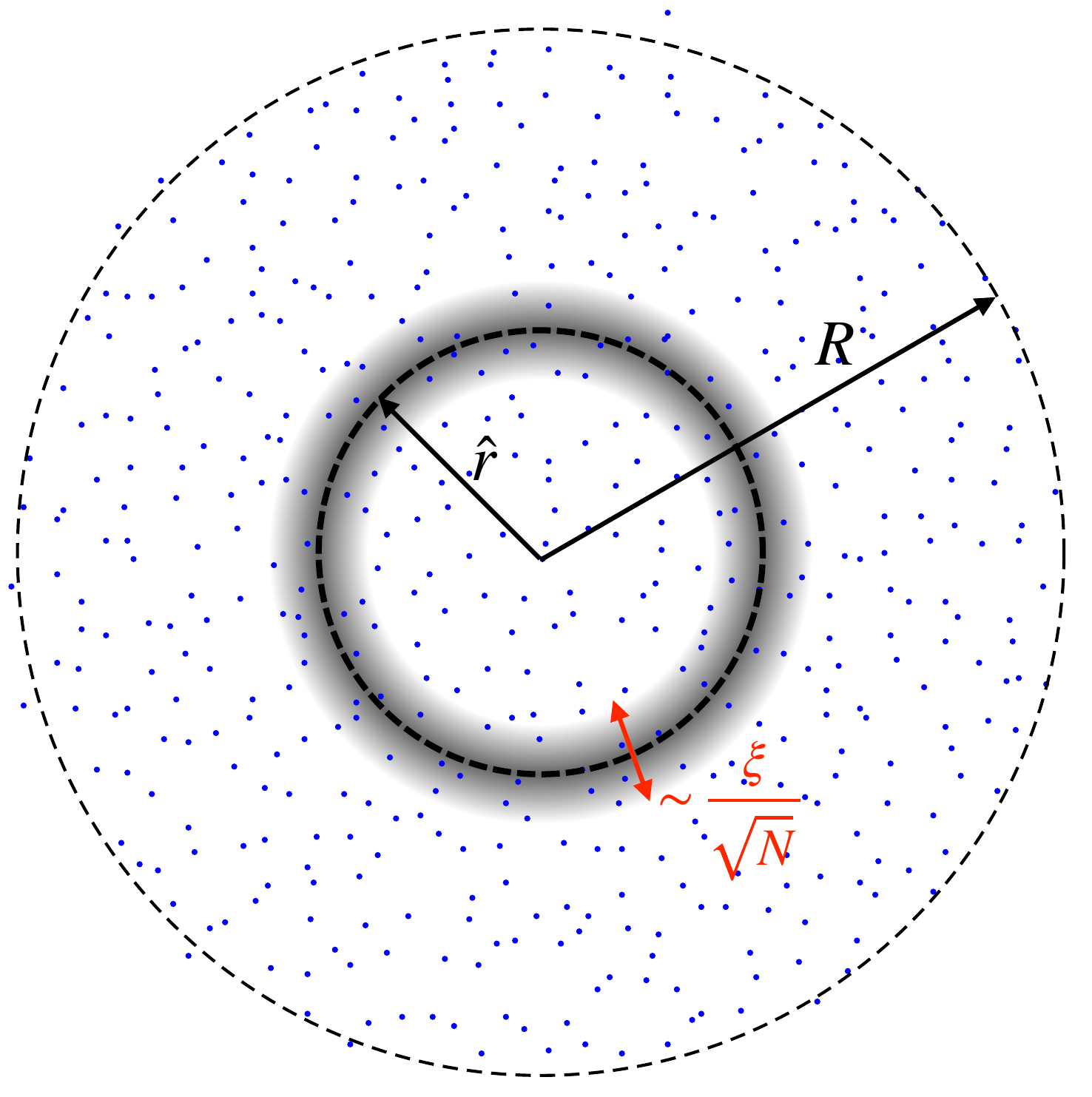}
\caption{Illustration of the {\it microscopic linear statistics} in Eq. (\ref{def_micro}) studied in this paper. Here, the blue points correspond to the eigenvalues of an $N \times N$ Ginibre matrices, with $N=500$, described by a Coulomb gas as in Eq. (\ref{PDF_intro}) with $U(r)=r^2/2$. In this case the equilibrium density has support on the disk of radius $R$, centered at the origin. The shaded area around the circle of radius $\hat r$ and width $\sim \xi/\sqrt{N}$
(depicted in gray scale) indicates the microscopic region (of size comparable to the interparticle spacing) where the linear statistics (\ref{def_micro}) takes nonzero values.}\label{Fig_intro}
\end{figure}

The aim of this paper is to analyze the case where the function $f(r)$ varies on a scale $\xi/\sqrt{N}$, i.e., of the order 
of the interparticle distance (see Fig. \ref{Fig_intro}). We will focus on normal complex random matrices (i.e., a 2D Coulomb gas with $\beta = 2$)
for which determinantal formulae allow us to compute the cumulants of ${\cal L}_N$. By varying
the scale $\xi$ we can probe the crossover from microscopic to smooth linear statistics
and investigate in more detail the peculiarity of the edge behavior. From the cumulant generating function, we also extract the large deviation form of the full distribution of ${\cal L}_N$, 
which until now was known only for smooth linear statistics \cite{BLMS23} in general space dimension,
and for the FCS in the 1D Coulomb gas 
\cite{Dhar_FCS,Flack_FCS,LSFCS}.

We now turn to the definition of the model and of the observables, as well as the presentation of the main results.

\section{Model, observables and main results}

\subsection{Model}

We consider the classical Coulomb gas in two spatial dimensions with $N$ particles at positions $\bx_1,\dots,\bx_N$ in the presence of a central potential $U(r = |\bx|)$. At canonical equilibrium with inverse temperature $\beta$ the joint probability distribution function (PDF) of the positions 
${\cal P}(\bx_1,\ldots,\bx_N)$
is given by
\begin{align} \label{Pjoint}
  {\cal P}(\bx_1,\ldots,\bx_N) = \frac{1}{Z_{N,U}} e^{- \beta \, {\cal E}(\bx_1,\ldots,\bx_N)} \;, 
\end{align}
where the total energy reads
\begin{align} \label{energy}
  {\cal E}(\bx_1,\ldots,\bx_N) = - \sum_{1\leq i<j\leq N} \log(|\bx_i-\bx_j|) + N \sum_{k=1}^N U(|\bx_k|)\,.
\end{align}
In Eq. (\ref{Pjoint}), $Z_{N,U}$ is the partition function defined as 
\bea \label{eq:Z}
Z_{N,U} = \int_{\mathbb{R}^2} d\bx_1 \cdots \int_{\mathbb{R}^2} d\bx_N \, e^{-\beta\left[-\sum_{1\leq i<j\leq N} \log (|\bx_i-\bx_j|) + N \sum_{k=1}^N U(|\bx_k|)\right]} \;.
\eea
For later purpose it is convenient to indicate the external potential as a subscript. 

In this paper we focus on the case $\beta=2$ which, as we discuss below, has a determinantal 
structure allowing for detailed analytical calculations. In that case,
the joint PDF in Eq.~\eqref{Pjoint} also
describes the joint PDF of the eigenvalues $\bx_i \to z_i \in \mathbb{C}$ of
an $N \times N$ so-called random normal matrix $M$, drawn from the probability density 
$\sim e^{- N {\rm Tr} [\, U(M)]}$ \cite{Zaboronsky98,Ameur2015}, where ${\rm Tr}(M)$ denotes the trace of the matrix $M$, which 
reads
\begin{align} \label{PDF_intro}
  P(z_1,\ldots,z_N) = \frac{1}{Z_{N,U}}\prod_{i<j}|z_i-z_j|^2 \prod_{i=1}^N e^{-2\,N U(|z|)}\,.
\end{align}
The particle positions in the complex plane thus coincide with the eigenvalues of $M$. Furthermore,
for the choice $U(r)=r^2/2$, Eq. (\ref{PDF_intro}) corresponds to the joint PDF of the eigenvalues $z_i \in \mathbb{C}$ of the complex Ginibre ensemble~\cite{Mehtabook,Forrester}.

The simplest case that we focus on is a potential $U(r)$ which is smooth and confining, so that the particles remain in the vicinity of the origin
\footnote{Here one assumes that $U(\bx) \gg \log |\bx|$ at large $|\bx|$, see e.g., \cite{saff}.}.
Note that in Eq. (\ref{energy}) the effective potential is actually $N\,U(\bx)$, such that the equilibrium average density of particles has a support of order $O(1)$ in the limit $N \to \infty$. More precisely the mean density defined as 
\be \label{mean_density} 
\bar \rho_N(\bx) = \left\langle \frac{1}{N} \sum_{i=1}^N \delta(\bx-\bx_i) \right\rangle_{U} \;,
\ee 
where $\langle \cdots \rangle_U$ denotes an average with respect to the Boltzmann weight in (\ref{Pjoint}), 
converges in the limit $N \to \infty$ to the equilibrium density given by (within its support, and denoting $r= |\bx|$)
\be \label{rho_eq}
\lim_{N \to \infty} \bar \rho_N(\bx) = \rho_{\rm eq}(r)= \frac{1}{2 \pi} \Delta U(r) = \frac{1}{2 \pi r} (r U'(r))'  = \frac{1}{2 \pi}\left(\frac{U'(r)}{r} + U''(r)\right)\;,
\ee
which is of order $O(1)$ as $N \to \infty$. 
Note that, here, we restrict to potentials $U(r)$ such that the support of $\rho_{\rm eq}(r)$ is a disk centered at the origin and of radius $R$. Note that for $r = R$ we 
define 
\be \label{rho_rmoins}
\rho_{\rm eq}(R)=\rho_{\rm eq}(R^-)>0  \;.
\ee 
The normalization condition determines the value of $R$ as the unique solution of 
\be \label{def_R}
\int d\bx \, \bar \rho_N(\bx) = 1 \quad \Longrightarrow \quad R\,U'(R)=1 \;.
\ee 
In addition one must assume that $r U'(r)$ is a strictly increasing function of $r \in [0,R]$ so that
the density is everywhere strictly positive in the disk.

\subsection{Linear statistics}

We are interested in the linear statistics, i.e., the fluctuations of ${\cal L}_N$ defined as
\begin{align}
  {\cal L}_N = \sum_{i=1}^N f_N(|\bx_i|) \,,\label{eq:lin}
\end{align}
where we allowed for an $N$-dependence in the linear statistics function. In many cases, it is
chosen to be $N$-independent, i.e. $f_N(r)=f(r)$ with either (i) $f(r)$ a smooth function 
or (ii) $f(r)$ has jumps such as $f(r)= \mathbb{I}_{0<r<\hat r}$ in which case
${\cal L}_N= {\cal N}_{\hat r}$ is the number of particles in a disk of radius $\hat r$ around
the origin. 

To interpolate between these situations in the present work we consider the form (see Fig. \ref{Fig_intro}) 
\be \label{def_micro}
  f_N(r) = \phi\left( \frac{(r-\hat r) \sqrt{N}}{\xi}  \right) \;,
\ee
where $0 < \hat r \leq R$ is a reference radius within the droplet,
$\phi(u)$ is a fixed smooth function and $\xi$ a parameter, with the interpretation of a characteristic length scale. 
Since the interparticle distance is of order $1/\sqrt{N}$ this linear statistics probes the microscopic 
scales. As $\xi \to +\infty$, we expect -- and will explicitly verify -- that this limit recovers the case (i) of smooth macroscopic linear statistics.

To study the distribution of the linear statistics in (\ref{eq:lin}), one defines its cumulant generating function (CGF).
There are two definitions which are convenient in two distinct scaling regimes at large $N$. The first one, denoted
$\chi(s,N)$, 
is used to obtain the cumulants for smooth linear statistics (i) above, as done in \cite{BLMS23}, 
\begin{align}
 \chi(s,N) =  \log \langle e^{- N s \, {\cal L}_N } \rangle_U = 
 \log \langle e^{- N s \sum_{i=1}^N f(\bx_i)}\rangle_U \,. \label{eq:chidef}
\end{align}
The second definition is useful to obtain the cumulants of the counting statistics {as in case (ii)} and reads \cite{Lacroix_rotating,castillo}
\begin{align}
 \tilde \chi(\mu,N) =  \log \langle e^{- \mu \, {\cal L}_N } \rangle_U = 
 \log \langle e^{- \mu \sum_{i=1}^N f(\bx_i)}\rangle_U \,. \label{eq:chidef}
\end{align}
Of course at finite $N$ the cumulants of ${\cal L}_N$, denoted as $\langle {\cal L}_N^q \rangle_c$, can be obtained 
from an expansion in
the parameter $s$ or $\mu$, 
using either definition
\be \label{cum_gen}
\chi(s,N) = 
\sum_{q \geq 1} \frac{(-1)^q}{q!} N^q s^q \langle {\cal L}_N^q \rangle_c \quad, \quad \tilde \chi(\mu,N) = \sum_{q \geq 1} \frac{(-1)^q}{q!} \mu^q \langle {\cal L}_N^q \rangle_c
\;,
\ee
where $\chi(s,N)=\tilde \chi(\mu=N s, N)$, 
where $s$ and $\mu$ are assumed to be sufficiently small to ensure the convergence of the series in Eq. (\ref{cum_gen}). 

For later purpose, we note that the first derivative of the CGF can also be written as~\cite{BLMS23} 
\be  \label{trick0} 
 \partial_s  \chi(s,N)  = - N^2 \int_0^R 2 \pi r dr \bar \rho_{s,N}(r) f(r) \;, 
\ee
where $\bar \rho_{s,N}(r)$ converges at large $N$ to the equilibrium density in the shifted potential $U + \frac{s}{2} f$,
i.e. from \eqref{rho_eq}, $\bar \rho_{s,N}(r) \to \rho_{s,{\rm eq}}=\frac{1}{2 \pi} \Delta (U + \frac{s}{2} f)$ as $N \to \infty$,
which below is called the ``constrained density''. 

\subsection{Main results} 

Let us now summarize our main results. To probe the microscopic scales we consider a linear statistics with a function $f_N(r)$ of the form \eqref{def_micro}
where $N$ is large but the length $\xi$ is fixed. To compare this 
length with the typical interparticle distance we write it as 
\be  \label{xidef} 
\xi = 
 \frac{1}{\sigma} \frac{1}{\sqrt{4 \pi  \rho_{\rm eq}(\hat r)} } \;,
\ee 
where $\rho_{\rm eq}(\hat r)$ 
was given in \eqref{rho_eq} and \eqref{rho_rmoins} for $\hat r \leq R$
and 
$\sigma$ is a free dimensionless parameter. 
Small values of $\sigma$ then correspond to
the limit towards macroscopic scales. 

We first obtain the following formula for the $q$-th order cumulant at the microscopic scale, to leading order for large~$N$
\begin{align} \label{cumintro} 
\langle {\cal L}^q_N \rangle_c \simeq  \sqrt{N \,  \pi \hat r^2 \rho_{\rm eq}(\hat r)} \frac{1}{\sigma}   \int_{-\infty}^{y_{\max}} dy \; \overline{\phi(y + v)^q}^c \quad, \quad y_{\max} = \begin{cases}
&+ \infty \;, \; \hat r < R \\
&0 \;, \; \quad \;\hat r = R \;,
\end{cases} 
\end{align}
where the overbar $\overline{F(v)}$ denotes the average of $F(v)$ over 
a centered Gaussian random variable $v$, of variance $\sigma^2$, and $\overline{ F(v)^q }^c$ denotes the corresponding $q$-th cumulants of $F(v)$. The above result is valid provided the integral is convergent. 
At the edge of the droplet, for $\hat r=R$, the integration upper bound must be set to zero. 
Note that for $\hat r>R$ the cumulants are zero to the order $\sqrt{N}$ (in fact one expects that they are
exponentially suppressed at large $N$).

As we explained in the introduction, one of the motivation of this work is to understand 
how one can match the dependence $\sim N^{2-q}$ of the cumulants for the smooth linear statistics,
with the $\sim \sqrt{N}$ dependence of the cumulants for the counting statistics. To investigate this crossover between the macroscopic and the microscopic scales
we perform an expansion at small $\sigma$ (i.e., for large $\xi$) of the microscopic result (to leading order in $N$) given
in~\eqref{cumintro}, which we denote here $\langle {\cal L}^q_N \rangle_c^{\rm micro}(\phi,\sigma)$.
We show that (up to higher order terms in~$\sigma$) 
\be \label{matchinggeneral} 
\langle {\cal L}^q_N \rangle_c^{\rm micro}(\phi(u),\sigma)   \underset{\sigma \to 0}{\simeq}
\langle {\cal L}^q_N \rangle_c^{\rm macro}(f(r))|_{f(r) \to \phi(\sqrt{N}(r- \hat r)/\xi)}   \;.
\ee 
The notation in the right hand side (r.h.s.) denotes the formula for smooth functions $f(r)$ obtained
in the macroscopic regime in 
\cite{BLMS23}, to leading order in $N$, where we have replaced $f(r)$ by its form \eqref{def_micro}. 
Since the macroscopic formula is valid only for smooth functions, this
replacement is a priori not justified. However we show that, nevertheless, the identity \eqref{matchinggeneral} holds,
both in the bulk $\hat r<R$ and at the edge $\hat r=R$. 
This shows the smooth matching (with no additional intermediate regime of length scale)
between the microscopic result where
the cumulants are $O(\sqrt{N})$, and the macroscopic one, where they are $O(N^{2-q})$. The extra powers of $N$ 
come from the substitution from $f$ to $\phi$,
in their derivatives.

A remarkable feature discovered in \cite{BLMS23} is that for cumulants of order $q \geq 3$ and in the bulk, the
r.h.s. of \eqref{matchinggeneral} is identically zero. Here we show how this actually occurs. 
Indeed although the microscopic result for these cumulants on the left hand side (l.h.s.), is always non zero for $\sigma>0$,
its leading term in the small $\sigma$ expansion vanishes for $\hat r<R$.
Instead, at the edge, for $\hat r=R$, this leading term does not vanish, as found in Ref. \cite{BLMS23}. In fact, in this case, our result (\ref{matchinggeneral}) shows that that the leading contribution to the cumulants of order $q \geq 3$ comes from a boundary layer of size $O(1/\sqrt{N})$ near the boundary of the droplet.

We have also studied the PDF, ${\cal P} ( {\cal L}_N  )$, of the linear statistics ${\cal L}_N$
at the microscopic scale associated to 
the function $\phi(u)$ in \eqref{def_micro}. It exhibits large deviation forms in the large $N$ limit with two different
regimes. The first one, where ${\cal L}_N \sim \sqrt{N}$, which originates from fluctuations
at the microscopic scales, and a second one, where ${\cal L}_N \sim N$, originating from
macroscopic fluctuations, also referred to as {\it the Coulomb gas regime}. Namely one has
\bea \label{summary_LDF}
{\cal P} ( {\cal L}_N  ) \simeq 
\begin{cases}
&\exp \left(  - \sqrt{N} \,\Psi\left(  w=\frac{  {\cal L}_N  }{\sqrt{N} } \right)   \right) \;, \quad \; {\cal L}_N = O(\sqrt{N}) \\
& \\
& \exp \left(  - N^2 \,\Psi_{\rm CG}\left(   \Lambda= \frac{  {\cal L}_N  }{N } \right)   \right) \;, \; \;\;{\cal L}_N = O(N) \;.
\end{cases}
\eea  
We have obtained the rate functions $\Psi(w)$, see Eqs. \eqref{Psi} and \eqref{param}, as well as $\Psi_{\rm CG}(\Lambda)$, in the case of the Ginibre ensemble, see \ref{sec:CG}. The function $\Psi(w)$
behaves quadratically close to its minimum, see 
\eqref{psi_gauss}, and exhibits a cubic behavior at large argument $w \to  +\infty$
\be \label{cubicintro}
\Psi(w) \simeq A w^3  \;,
\ee 
where the amplitude $A$ is given in \eqref{Apm}. 
The rate function $\Psi_{\rm CG}(\Lambda)$ also 
exhibits a cubic behavior but at small $\Lambda$. We
find indeed
$\Psi_{\rm CG}(\Lambda) \simeq  A  \Lambda^3$,
with the same amplitude $A$. This shows the perfect
matching between the two regimes in (\ref{summary_LDF}). 

This cubic behavior differs significantly from the 
quadratic behavior of $\Psi_{\rm CG}(\Lambda)$ obtained in \cite{BLMS23}
in the case of smooth macroscopic linear statistics. It arises here
because the function $\phi(u)$ probes a microscopic region of the system.
To obtain values of ${\cal L}_N$ of order $N$, i.e. much 
larger than ${\cal L}_N \sim \sqrt{N}$, one needs to concentrate 
a finite fraction of the charges near $\hat r$ (corresponding to the delta function in Fig. \ref{Fig_hole}), resulting
in the formation of a macroscopic hole, see Fig. \ref{Fig_hole}.
A related phenomenon was observed in the case of the counting 
statistics~\cite{allez,castillo}. Our calculation provides a clearer insight into the characteristic scales involved in this process.

\subsection{Outline}

The rest of the paper is organized as follows. In Section \ref{sec:exact}
we display the exact formula for the cumulant generating function, valid for any $N$
and for an arbitrary function $f(r)$. 
In Section \ref{sec:cumul} we study various limits of this formula at large $N$.
In subsection \ref{subsec:previous} we recall the results obtained
in \cite{BLMS23} in the case of a smooth
linear statistics. The special case of a linear function $f(r)$ is discussed in \ref{app:linear}. 
In subsection \ref{subsec:cumul} we obtain the formula for the generating function and for the cumulants
in the case of the microscopic linear 
statistics defined in \eqref{def_micro}. We give a few examples of
applications, which include the counting statistics inside an annulus.
In subsection \ref{subsec:matching} we show how our microscopic result
matches the one for smooth linear statistics in the limit $\xi \to \infty$ (i.e., $\sigma \to 0$).
More details of these calculations are given in \ref{app_cumul}. In Section \ref{sec:ldf} we study the large deviation form of the
PDF of ${\cal L}_N$ for the microscopic linear statistics. 
We obtain the two rate functions in the regimes ${\cal L}_N \sim \sqrt{N}$
and ${\cal L}_N \sim N$ respectively, and show how they smoothly match. The shape of
the density conditioned to these large values for ${\cal L}_N$ is discussed. It is computed in details,
together with the rate function $\Psi_{\rm CG}(\Lambda)$, in
\ref{App:const_density} and \ref{sec:CG}.


\section{Exact determinantal formula for the cumulant generating function} \label{sec:exact}

For the normal random matrix ensemble with $\beta=2$ the
eigenvalues form a determinantal point process, and the joint PDF in \eqref{Pjoint}
takes the form
\begin{align}
P(z_1,\ldots,z_N) = \frac{1}{N!} \det_{1 \leq i,j \leq N} K_N(z_i, z_j) \;,
\end{align}
with the kernel
\be 
K_N(z,z') = \sum_{\ell=0}^{N-1} \frac{(z \bar z')^\ell}{h_{\ell,N}}   e^{- N U(|z|) - N U(|z'|)} \;,
\ee 
where
\be 
h_{\ell,N} = 2 \pi \int_0^{+\infty} dr r^{2 \ell+1} e^{- 2 N U(r) } \;.
\ee 
In particular, the exact mean density for any $N$ is given by 
\be \label{def_rho}
\overline{\rho}_N(|z|) = \frac{1}{N} K_N(z,z) = \frac{1}{N} \sum_{\ell=0}^{N-1} \frac{|z|^{2\ell}}{h_{\ell,N}}   e^{- 2 N U(|z|) } 
\ee 

To express the CGF we can use the standard determinantal formula 
\begin{align} 
&&\tilde \chi(\mu,N) = \log \langle \prod_{i=1}^N  e^{-\mu f_N(|z_i|} \rangle 
= \log {\rm Det}\left(I - ( 1- e^{-\mu f_N} ) K_N) \right) \label{logdet1}\\
&&= \log {\rm det}_{0 \leq \ell,\ell' \leq N-1} (I_N - A_{\ell,\ell'} ) \;, \label{logdet2}
\end{align}
where, in the second line, we have used the so called Sylvester identity. {Note that in Eq. (\ref{logdet1}) the notation ``${\rm Det}$'' means a Fredholm determinant, while in (\ref{logdet1}) the notation ``${\rm det}$'' stands for a standard determinant of an $N \times N$ matrix. Besides, in Eq. (\ref{logdet2}), the overlap matrix reads} 
\begin{align}
A_{\ell,\ell'} = \int d^2 z \frac{z^\ell \bar z^{\ell'} }{\sqrt{h_{\ell,N} h_{\ell',N}  }} e^{- 2 N U(|z|)- \mu f_N(|z|)} 
= \delta_{\ell,\ell'} \frac{2 \pi}{h_{N,\ell}} \int_0^{+\infty} dr \, r^{2 \ell+1}  e^{-2 N U(r)  - \mu  f_N(r)} \;,
\end{align}
where we have used the rotational invariance of the potential $U \equiv U(|z|)$ and of the linear statistics $f_N(z) \equiv f_N(|z|)$ to perform the integral over the angular variable, leading to $A_{\ell,\ell'} \propto \delta_{\ell,\ell'}$. Since the  overlap matrix is diagonal,  
the determinant is readily evaluated and one obtains the CGF $\tilde \chi(\mu,N)$ for arbitrary $U(r)$, $f_N(r)$ and $N$ as
\begin{align} \label{det_form}
\tilde \chi(\mu,N) = \sum_{\ell=0}^{N-1} \log  \left( \frac{ \int_0^{+\infty} dr \, r^{2 \ell+1}  e^{-2 N U(r)  - \mu f_N(r)} }{\int_0^{+\infty} dr r^{2 \ell+1}  e^{-2 N U(r)} } \right) \;.
\end{align}
We recall that the case of the complex Ginibre ensemble corresponds to $U(r)=r^2/2$. 


\section{Formula for the cumulants at large $N$}\label{sec:cumul}

\subsection{A reminder of previous results for smooth linear statistics} \label{subsec:previous}

Here we start by briefly recalling the method and the results in the case $f_N(r)=f(r)$ of smooth
linear statistics, as obtained in \cite{BLMS23}. In that case one sets $\mu = N s$
and study $\chi(s,N)$ as defined in \eqref{eq:chidef}. In the large 
$N$ limit one can write $\ell = \lambda N$ and approximate
the sum over $\ell$ in \eqref{det_form} by an integral, leading to
\be \label{chiI} 
 \chi(s,N) \simeq N \int_0^1d\lambda \log \left(\frac{  I_N(s,\lambda) }{I_N(0,\lambda)}\right)  \quad 
, \quad I_N(s,\lambda)=  \int_0^{+\infty} dr \, r\,  
e^{- N ( 2 U(r) - 2 \lambda \log r - s f(r) ) } \;.
\ee 
The integral $I_N(s,\lambda)$ can be evaluated by a saddle point method
and one obtains \cite{BLMS23}
\begin{align} \label{logI} 
    \log \left(\frac{  I_N(s,\lambda) }{I_N(0,\lambda)} \right)\simeq N \, \min_{r \geq 0} \phi_{s,\lambda}(r) = N \phi_{s,\lambda}(r_{s,\lambda}) \;,
\end{align}
where 
\begin{align}
 \phi_{s,\lambda}(r)  = 2 U(r)  - 2 \lambda \log r + s  f(r) - b(\lambda)  \label{phi} \;,
\end{align}  
and $b(\lambda)=\min_{r>0} ( 2 U(r) - 2 \lambda \log r)$. In Eq. (\ref{logI})
$r_{s,\lambda}>0$ minimises the function $\phi_{s,\lambda}(r)$, i.e., it is the solution of
\begin{eqnarray}
\partial_r \phi_{s,\lambda}(r)|_{r=r_{s,\lambda}} = 0 \quad \Longleftrightarrow \quad 
r_{s,\lambda}\, U'(r_{s,\lambda}) + \frac{s}{2} r_{s,\lambda} f'(r_{s,\lambda}) = \lambda \;. \label{saddle}
\end{eqnarray}
We note that $r_{s=0,\lambda=1}=R$, where $R$ is the radius of the droplet determined by (\ref{def_R}).
Since we have assumed that $r \, U'(r)$
is a monotonically increasing function on $r \in [0,R]$, it guarantees unicity of the solution of \eqref{saddle}
for $|s|$ sufficiently small. 

Inserting the asymptotics \eqref{logI} into \eqref{chiI} we obtain the following estimate for the
CGF to leading order at large $N$
\be \label{chi_largeN}
\chi(s,N)  \simeq N^2 \mathcal{F}(s) \;,
\ee 
with 
\be
   \mathcal{F}(s)  = - \int_0^1 d\lambda\,\min_{r\geq 0}[\phi_{s,\lambda}(r)] = - \int_0^1 d\lambda \, \phi_{s,\lambda}(r_{s,\lambda}) \,,\label{eq:F}
\ee 
Taking one derivative with respect to $s$ in Eq. (\ref{eq:F}), using $\partial_r \phi_{s,\lambda}(r)|_{r=r_{s,\lambda}}=0$ together with 
the explicit dependence of $\phi_{s,\lambda}(r)$ on $s$ in~\eqref{phi}, one finds
\begin{align} 
\partial_s  \mathcal{F}(s) = - \int_0^1 d\lambda f(r_{s,\lambda}) = 
- \int_0^{R_s} d\left( r_{s,\lambda}\, U'(r_{s,\lambda}) + \frac{s}{2} r_{s,\lambda} f'(r_{s,\lambda}) \right) f(r_{s,\lambda})
\label{Fprime} \;,
\end{align}
where in the last equation we have performed the change of variable $\lambda \to r_{s,\lambda}$, 
and expressed $d\lambda$ using \eqref{saddle}. Here we have defined $R_s$ as 
\be 
R_s = r_{s,\lambda=1} \quad \Longleftrightarrow \quad R_s U'(R_s) + \frac{s}{2} R_s f'(R_s) = 1  \;.\label{Rs2} 
\ee 
In conclusion, one obtains~\cite{BLMS23}
\be
\frac{1}{N^2} \partial_s \chi(s,N) \simeq \mathcal{F}(s) = -\int_0^{R_s} dr \left(r U'(r) + \frac{s}{2} r\,f'(r)\right)' f(r)\,,\label{eq:G2}
\ee
where $R_s$ is the solution of \eqref{Rs2}. 

\vspace*{0.5cm}
\noindent{\bf Remark 1}. The above method assumes that the function
\be 
\lambda(r) : = 
r U'(r) + \frac{s}{2} r  f'(r)   \quad , \quad \lambda'(r) = 2 \pi r \rho_{s,\rm eq}(r) 
\ee 
is a strictly increasing function of $r \in [0,R]$, so that the equation
\eqref{saddle} has a unique solution for all $\lambda \in [0,1]$.
This is equivalent to the ``constrained'' density $\rho_{s,\rm eq}(r)$ 
(associated to the shifted potential $U \to U+ (s/2 ) f$) being 
strictly positive for $r<R$.
In general this holds only for $|s|<s_c$, where a double root appears at $|s|=s_c$.
Indeed, exactly at $|s|=s_c$, there exists a value $r=r_c$ such that 
$\lambda'(r_c)= 0$, i.e. the constrained density vanishes. For $|s|>s_c$
the constrained density develops a multiple support, and the calculation
must be reconsidered. This is well known in the case of the counting
statistics for the Coulomb gas \cite{allez,castillo,Flack_FCS,Dhar2018}, where
an additional delta peak is generated in the density, 
as well as for the log-gas \cite{MNSV2009,MNSV2011,MV2012,MMSV2016,valov}. We will come back to
this remark below, in Section \ref{sec:ldf}, where this feature plays an important role
in the asymptotics of the rate functions associated to the large deviations of ${\cal L}_N$.
\\

Proceeding from there, taking \eqref{Fprime} at $s=0$, and using Eqs. (\ref{cum_gen}) and (\ref{chi_largeN}), one obtains the first cumulant
\begin{align} \label{first_cumul}
\langle {\cal L}_N\rangle \simeq - N{\cal F}'(0) = N \int_0^R (r\,U'(r))'\, f(r) \, dr  = N \int_0^R 2 \pi r \rho_{\rm eq}(r) f(r)\, dr\;,
\end{align}
where, in the last equality, we have used the expression of the equilibrium density 
from Eq. (\ref{rho_eq}).

Next, taking derivatives with respect to (w.r.t.) $s$ of \eqref{eq:G2} 
one obtains after some manipulations (see Ref. \cite{BLMS23}) the following expressions for the second and 
third 
cumulants in the large $N$ limit
\begin{align} \label{secondcum2} 
   &  \langle {\cal L}_N^2 \rangle_c \simeq \frac{1}{2} \int_0^{R} dr \, r [f'(r)]^2  \;, \\
    & \langle {\cal L}_N^3 \rangle_c \simeq \frac{1}{4 N} 
\frac{R f'\left(R\right)^3}{
   U''\left(R\right)+\frac{U'(R)}{R}} = \frac{1}{8 \pi N} \frac{Rf'(R)^3}{\rho_{\rm eq}(R)}  \;, \label{thirdcum2}
\end{align}
where we recall that the radius $R$ is such that $U'(R) R = 1$, and we denote $\rho_{\rm eq}(R) \equiv \rho_{\rm eq}(R^-)$
the density at the edge of the droplet (which jumps to zero outside).
We can also obtain 
the following result for the general cumulant of order $q \geq 3$ \cite{BLMS23}
\be \label{cumul_macro}
\langle {\cal L}_N^q \rangle_c \simeq \frac{1}{2^{q-1} N^{q-2}} \, (A(r) \partial_r)^{q-3}  \left( A(r) r f'(r)^2  \right)|_{r=R} 
\quad , \quad U'(R) R = 1 \;,
\ee 
with 
  \be  \label{def_A}
A(r) = 
\frac{ f'\left(r\right){}^2}{
   f''\left(r\right) \left(r^{-1} -
   U'\left(r\right)\right)+f'\left(r\right)
   \left( U''\left(r\right)+\frac{1}{r^2} \right)} \;.
   \ee 
Although these results are given here for $\beta=2$, in Ref. \cite{BLMS23} we have shown that the $\beta$-dependence of the cumulant of order $q$
is obtained from (\ref{cumul_macro}) by substituting $2^{q-1}$ by $\beta^{q-1}$, hence $\langle {\cal L}_N^q \rangle_c  \propto \beta^{1-q}$. 



\subsection{Cumulants of the linear statistics at microscopic scale for large $N$} \label{subsec:cumul}

\subsubsection{General formula} 

We now study a completely different regime, where the linear statistics function is of the form 
\be \label{def_micro2}
  f_N(r) = \phi\left( \frac{(r-\hat r) \sqrt{N}}{\xi}  \right) \;.
\ee
We start by assuming that $\phi(u)$ is a smooth test function, which is bounded and varies on a scale $u=O(1)$. Let us start with the large $N$ formula \eqref{chiI}, setting $s = \mu/N$, namely 
\bea  
&& \tilde \chi(\mu,N) 
\simeq N \int_0^1 d\lambda \, \log \left(\frac{  \tilde I_N(\mu,\lambda) }{\tilde I_N(0,\lambda)} \right)\label{chitilde_mu} \\
&& \tilde I_N(\mu,\lambda)=  \int_0^{+\infty} dr \, r\,  
e^{- N ( 2 U(r) - 2 \lambda \log r ) - \mu  \phi\left( \frac{(r-\hat r) \sqrt{N}}{\xi}  \right)  } \;, \label{Itilde} 
\eea  
where the only hypothesis we made is to assume that the sum over $\ell$ in \eqref{det_form} is dominated by large values
of $\ell$. 

Since the function $\phi(u)$ varies on a scale $u=O(1)$, the logarithm of the ratio in Eq. \eqref{chitilde_mu} is dominated by the region of $r$ near $\hat r$.
Hence we will perform the following change of variable
\be \label{def_hatlambda}
r = \hat r + \xi \frac{u}{\sqrt{N}} \quad , \quad \lambda = \hat r U'(\hat r) + \frac{\tilde \lambda}{2 \xi \sqrt{N}} \hat r \;.
\ee 
Expanding the term in the exponential in Eq. (\ref{Itilde}) at large $N$ one finds
\bea  \label{expansion1}
&& 2 N ( U(r) -  \lambda \log r) = 2 N (U(\hat r) - \hat r U'(\hat r) \log \hat r) -  \sqrt{N} \tilde \lambda \frac{\hat r}{\xi}  \log \hat r \\
&& - \tilde \lambda u  + \xi^2 u^2 \left( \frac{U'(\hat r)}{\hat r} + U''(\hat r)\right) 
+ \frac{1}{\sqrt{N}} P_3(u, \tilde \lambda) + \frac{1}{N} P_4(u, \tilde \lambda) + o\left(\frac{1}{N}\right) \;,\nonumber 
\eea  
where $P_3$ and $P_4$ are homogeneous polynomials in $u$ and $\tilde \lambda$ of degrees three and four respectively. For convenience we will set
\be  \label{xidef} 
\xi^2 = \frac{1}{2 \sigma^2 \left(\frac{U'(\hat r)}{\hat r} + U''(\hat r)\right)} = \frac{1}{4 \pi \sigma^2 \rho_{\rm eq}(\hat r)} \;,
\ee 
where $\sigma$ is a free dimensionless parameter. At this stage we need to consider two different cases:

\begin{itemize}

\item[(i)]{The bulk region, $0 < \hat r < R$: from Eqs. (\ref{chitilde_mu}), (\ref{Itilde}) and (\ref{expansion1}), we obtain, discarding the subdominant terms at large $N$  
\be \label{chi_tilde_bulk}
\tilde \chi(\mu,N) \simeq \sqrt{N} \frac{\hat r}{2 \xi} \int_{-\infty}^{+\infty} d\tilde \lambda
\log \frac{  J(\mu,\tilde \lambda) }{J(0,\tilde \lambda)}  
~,~ J(\mu,\tilde \lambda)=  \int_{-\infty}^{+\infty} du
\, e^{   \tilde \lambda u - \frac{u^2}{2 \sigma^2}   - \mu \, \phi(u) } \;.
\ee 
}
\item[(ii)] The edge region, $\hat r = R$: In that case, since the integral over $\lambda$ in Eq.~(\ref{chitilde_mu}) is over 
$\lambda \in [0,1]$ and since $R U'(R)=1$ we see that the upper bound
on $\tilde \lambda$ is $0$ and one obtains instead
\be \label{chi_tilde_edge}
\tilde \chi(\mu,N) \simeq \sqrt{N} \frac{\hat r}{2 \xi} \int_{-\infty}^{0} d\tilde \lambda
\log \left(\frac{  J(\mu,\tilde \lambda) }{J(0,\tilde \lambda)}\right)  \;.
\ee 
\end{itemize}
We note that the CGF is well defined for any $\mu$, and scales as $\tilde \chi(\mu,N) \sim \sqrt{N}$
in this microscopic regime. This behavior is quite different from 
the scaling $\chi(s,N) \sim N^2$ for smooth macroscopic linear statistics -- see Eq. (\ref{chi_largeN}).


From Eqs. (\ref{chi_tilde_bulk}) and (\ref{chi_tilde_edge}), we can now obtain the cumulants by taking derivatives w.r.t. $\mu$ as in \eqref{cum_gen}.
They read 
\be \label{cumul_lmax}
\langle {\cal L}^q_N \rangle_c \simeq \sqrt{N} \frac{\hat r}{2 \xi}  \int_{-\infty}^{\tilde \lambda_{\rm max}} d\tilde \lambda (- \partial_\mu)^q 
\log  J(\mu,\tilde \lambda) |_{\mu=0} \;,
\ee 
where we denote the upper bound of the integral as
\be \label{def_lmax}
\tilde \lambda_{\rm max} = \begin{cases}  
 +\infty\;, \quad \hat r < R \qquad \rm{(bulk)}\\
0 \;, \quad \quad \,\hat r = R  \qquad \,\rm{(edge)}
\end{cases} \;.
\ee 
The formula \eqref{cumul_lmax} holds provided the corresponding integral over $\tilde \lambda$ is convergent (see the discussion below).  
It is convenient to perform the change of variable $u =  \sigma^2 \tilde \lambda + v$ in the integral over $u$ in $J(\mu,\tilde \lambda)$ in Eq. (\ref{chi_tilde_bulk}) and rewrite it as  
\be 
J(\mu,\tilde \lambda) =   e^{ \frac{1}{2}\sigma^2 \tilde \lambda^2} \int_{-\infty}^{+\infty} dv
\, e^{ - \frac{v^2}{2 \sigma^2}   - \mu \phi(v +  \sigma^2 \tilde\lambda ) } \;.
\ee 
This leads to the following formula for the cumulants
\begin{align} 
\langle {\cal L}^q_N \rangle &\simeq \sqrt{N} \frac{\hat r}{2 \xi}   \int_{-\infty}^{\tilde \lambda_{\rm max}} d\tilde \lambda \, 
\overline{ \phi( \sigma^2 \tilde \lambda + v)^q  }^c \nonumber \\
& \label{gen_cumul} =  \sqrt{N} \frac{\hat r}{2 \sigma^2 \xi}   \int_{-\infty}^{y_{\max}} dy \; \overline{\phi(y + v)^q}^c \quad, \quad y_{\max} = \begin{cases}
&+ \infty \;, \; \hat r < R \\
&0 \;, \; \quad \;\hat r = R \;,
\end{cases} 
\end{align}
where we performed the change of variable $y=  \sigma^2 \tilde \lambda$ in the last equality, and we used the overbar to denote the average over 
an independent Gaussian random variable of variance $\sigma^2$, i.e. one has
\be \label{Gauss}
\overline{ F(v) } =   \int_{-\infty}^{+\infty} \frac{dv}{\sigma \sqrt{2 \pi} } F(v) \, e^{  - \frac{v^2}{2 \sigma^2}  } \;,
\ee 
and $\overline{ F(v) }^c$ denotes the corresponding cumulants. Note that these cumulants can also be computed from the CGF as
\begin{align}  \label{formula_cumul}
& \langle {\cal L}^q_N \rangle_c = (- \partial_\mu)^q \tilde \chi(\mu, N) \Big\vert_{\mu=0} \quad, \quad 
\tilde \chi(\mu, N) \simeq \sqrt{N} \frac{\hat r}{2 \xi {\sigma^2}} \int_{-\infty}^{y_{\max}} dy \; \log \left(\overline{ e^{- \mu \, \phi(y + v) }} \right) \;.
\end{align} 

\vspace*{0.3cm}
\noindent Several remarks are in order:
\vspace*{0.3cm}

\noindent {\bf Remark 2 (constrained density)}. Taking a derivative w.r.t. $\mu$ of \eqref{chi_tilde_bulk}, one obtains the microscopic analog of the formula \eqref{trick0},
which reads 
\bea \label{dchitilde_txt}
\partial_\mu \tilde \chi(\mu,N) \simeq -\sqrt{N} 2 \pi \, \hat r \xi \int_{-\infty}^{+\infty} du \, \rho_\mu(u) \phi(u) \;. 
\eea
where $\rho_\mu(u)$ is the scaling form of the constrained density in the shifted potential 
\begin{align} \label{def_rhomu}
\overline{\rho}_{N}(r)\Big \vert_{U \to U + \frac{\mu}{2 N} \phi } &\simeq \rho_\mu \left(\sqrt{N}\frac{r-\hat r}{\xi} \right) \\
\rho_\mu(u) &= \frac{\sigma^2}{\pi} e^{-\frac{u^2}{2 \sigma^2}- \mu \phi(u)} \int_{-\infty}^{+\infty} d\tilde \lambda \frac{e^{\tilde \lambda u}}{\int_{-\infty}^{+\infty} du \, e^{-\frac{u^2}{2 \sigma^2}+\tilde \lambda u - \mu \phi(u)}} \;.
\end{align}
One can check that this formula is consistent with the formula given for $\tilde \chi(\mu,N)$ in Eq.~(\ref{chi_tilde_ratio}).

\noindent {\bf Remark 3}.
A formula similar to \eqref{gen_cumul} for the special case of the counting statistics was obtained in Ref. \cite{Lambert2022} (see Eq. (2.7) there).

\noindent{\bf Remark 4}. A heuristic interpretation of the result in Eqs.  \eqref{gen_cumul} and \eqref{formula_cumul},
is that 
the linear statistics
$ {\cal L}_N  := \sum_i \phi(u_i)$, where $u_i = (r_i - \hat r) \sqrt{N}/\xi$, 
behaves as the following sum of random variables (i.e. an integral in the large $N$ limit)
${\cal L}_N  \overset{\rm law}{=}  \int du \phi(u+v(u))$ where $v(u)$ are independent and identically distributed (i.i.d.) Gaussian random variables. 



\vspace*{0.5cm}
\noindent{\bf Remark 5}. As shown in \cite{kostlan} and discussed furhter in \cite{castillo} the joint distribution of the radii $r_1 = |z_1|, r_2 = |z_2|, \cdots, r_N = |z_N|$ takes the
following form 
\bea \label{Prad}
\hspace*{-2cm}P_{\rm rad}(r_1, r_2, \cdots, r_N) = \frac{1}{N!} \sum_{\sigma \in {\cal S}_N} \prod_{k=1}^N \frac{r_k^{2 \sigma(k)-1}} {h_{\sigma(k)}}\, e^{-2 N U(r_k)} \;, \; h_k = \int_0^\infty r^{2k-1} e^{-2 N U(r)}\, dr \;,
\eea
where ${\cal S}_N$ is the group of permutations of $N$ elements. Hence, up to
a random permutation, they are distributed as independent but {\it non-identical} 
random variables, allowing for an alternative derivation for the above formula for the CGF in Eq. (\ref{det_form}).

\subsubsection{Some applications.}  

Let us now discuss some applications of these results. We start with the first moment which reads
\bea  \label{firstcumphi}
&& \langle {\cal L}_N \rangle \simeq 
 \sqrt{N} \frac{\hat r}{2 \sigma^2 \xi}  \int_{-\infty}^{y_{\max}} dy \, \overline{\phi(y + v)} \;, 
\eea   
provided the integral over $y$ is finite. This requires the function $\phi(y)$ to decrease
fast enough at infinity. This excludes, e.g., the case $\phi(+\infty)=1$ which arises 
for instance in the counting statistics in the disk of radius $\hat r$
centered at the origin, which corresponds to the choice $\phi(y)=\theta(-y)$. 
Indeed, in that case one knows that $\langle {\cal L}_N \rangle \sim N$, instead of $\langle {\cal L}_N \rangle \sim \sqrt{N}$ in (\ref{firstcumphi}). 

The second cumulant reads 
\bea  \label{second_cum}
&& \langle {\cal L}^2_N \rangle_c \simeq 
 \sqrt{N} \frac{\hat r}{2 \sigma^2 \xi}   \int_{-\infty}^{y_{\max}} dy \, \left( \overline{ \phi(y + v)^2}    - \left(  \overline{ \phi(y + v) } \right)^2  \right) \;, 
\eea   
provided the integral is finite. Let us point out that we can relax the condition that $\phi(u)$ is a smooth
function provided this integral is finite (this holds also for the higher cumulants).
Let us now give several examples.  

\vspace*{0.5cm}
\noindent{(i) {\bf Number of particles inside a disk of radius $\hat r$: $\phi(y)=\theta(-y)$}. 
Although the first cumulant is not of order $\sqrt{N}$, as mentioned above, the integral in (\ref{second_cum}) is finite even for the counting statistics $\phi(y)=\theta(-y)$.
This leads to
\bea  
\langle {\cal L}^2_N \rangle_c &\simeq& 
 \sqrt{N} \frac{\hat r}{2 \sigma^2 \xi}   \int_{-\infty}^{y_{\max}} dy \, {\rm Prob.}(y<-v) \; {\rm Prob.}(y>-v)  \label{v63} \\
&=& \sqrt{N} \frac{\hat r}{2 \sigma \xi}   \int_{-\infty}^{y_{\max}}  dy \, \frac{1}{4} {\rm erfc}\left( - \frac{y}{\sqrt{2}} \right) 
{\rm erfc}\left(  \frac{y}{\sqrt{2}} \right) \label{erfc}\\
&=& 
\sqrt{N} \hat r \sqrt{\rho_{\rm eq}(\hat r)} \times 
\begin{cases} 
& 1 \quad, \quad \quad \hat r < R\\
& 1/2  \quad, \quad  \hat r = R  \;,
\end{cases} 
\eea   
where $\rho_{\rm eq}(\hat r)$ is given in \eqref{rho_eq}. We recall that, in (\ref{v63}), $v$ denotes a Gaussian random variable of zero mean and variance unity, while in Eq. (\ref{erfc}) ${\rm erfc}(z) = 2/\sqrt{\pi}\int_z^{\infty} e^{-u^2}\, du$ is the complementary error function. In the 
case of the Ginibre ensemble for which $U(r) = r^2/2$ and $\rho_{\rm eq}(\hat r)= 1/\pi$, 
this result is in agreement with \cite{Lacroix_rotating}. 
Note that there is a crossover near the edge from the value $1$ to $0$ which can
be studied by considering $\phi(y)=\theta(y_0-y)$ and varying $y_0 \in (-\infty,+\infty)$,
see \cite{Lacroix_rotating}, the case $\hat r= R$ corresponding to $y_0=0$. 
This example shows that the fluctuations in the number of particles in a disk are, to leading order, of microscopic origin
in a shell of width $1/\sqrt{N}$ near the boundary of the disk. 

Moreover it is easy to see that one can apply the general formula for the cumulants $\langle {\cal L}^q_N \rangle_c$ in Eq. (\ref{gen_cumul})
to $\phi(y)=\theta(-y)$ and recover, for any $q \geq 2$ the result for the cumulants of the counting statistics
obtained in \cite{Lacroix_rotating}. Actually, the simplest way to perform the calculation is to compute the generating function
in \eqref{formula_cumul} by substracting the first moment, as in \cite{Lacroix_rotating}. 

\vspace*{0.5cm}
\noindent{\bf Remark 6}. Let us consider the case of the bulk $\hat r<R$, i.e. $y_{\rm max}=+\infty$.
Performing the change $y \to -y$ and $v \to -v$ in the integral in \eqref{gen_cumul} 
we see that the 
leading term in the cumulants $\langle {\cal L}^q_N \rangle$ is identical for the function $\phi$ and the function $\tilde \phi$ defined
as $\tilde \phi(y)=\phi(-y)$. This is particularly useful in the case of the counting statistics discussed below
since for $\phi(y)=\theta(-y)$ one has $\tilde \phi(y)=1- \phi(y)$. This immediately shows that the $O(\sqrt{N})$ 
coefficient of
the odd cumulants vanish for $q \geq 3$. This property was already noticed in \cite{Lacroix_rotating} -- see also \cite{Charles}.

\vspace*{0.5cm}
\noindent{(ii) {\bf Number of particles in an annulus}.
Consider now an annulus of microscopic width, $r \in [ \hat r + \frac{y_1}{\sqrt{N}} \xi , \hat r + \frac{y_2}{\sqrt{N}} \xi]$
where $\xi$ is parametrized by $\sigma$ according to \eqref{xidef}. This corresponds to the choice
$\phi(y) = \theta(y_1 < y < y_2)$. Then from \eqref{second_cum} the variance reads
\bea  
&& \tilde \chi(\mu, N) \simeq \sqrt{N} \frac{\hat r}{2 \xi  \sigma^2} \int_{-\infty}^{y_{\max}} dy \; \log 
\left( 1 + (e^{-\mu}-1) B(y) \right) \;, \\
&&  B(y) = \frac{1}{2}  \left( {\rm erf}\left(\frac{y_2-y}{\sqrt{2 \sigma}  } \right)
- {\rm erf}\left(\frac{y_1-y}{\sqrt{2 \sigma}  } \right) \right) \;,
\eea  
where ${\rm erf}(z) = 1-{\rm erfc}(z) = 2/\sqrt{\pi} \int_0^z e^{-u^2}\,du$ is the error function. 
One finds
\bea 
\langle {\cal L}^2_N \rangle \simeq 
 \sqrt{N} \frac{\hat r}{2 \sigma^2 \xi} \int_{-\infty}^{y_{\max}} dy \; B(y) (1- B(y) ) \;.
\eea 
To compute this integral we rescale by $\sigma$, then we take a derivative w.r.t. $y_1$ and use the identity 
$\int_{-\infty}^\infty dx G'(x) G(a+b x) = G(a/\sqrt{1+b^2})$ where $G(x)=\int_{-\infty}^x \frac{dy}{\sqrt{2 \pi}} e^{-y^2/2}$,
see the formula 10.010.8 in \cite{error}. We obtain in the case $\hat r< R$, with $y_{21}=y_2-y_1$
\bea 
\langle {\cal L}^2_N \rangle \simeq 
 \sqrt{N} \frac{\hat r}{2 \sigma \xi} \left( \frac{2}{\sqrt{\pi}} (1- e^{- \frac{y_{21}^2}{4 \sigma^2} }) 
 + \frac{y_{21}}{\sigma} (2- {\rm erfc}\left(- \frac{y_{21}}{2\sigma} \right) \right) \;.
\eea 
One finds the following asymptotic behavior at small and large $y_{21}$ 
\be
\langle {\cal L}^2_N \rangle \simeq 
\begin{cases}
& \sqrt{N} \frac{\hat r}{2 \sigma^2 \xi} y_{21} (1 - \frac{y_{21}}{2 \sigma \sqrt{\pi} }+ O(y_{21}^2) )  \quad , \quad y_{21}/\sigma \ll 1  \\
& \sqrt{N}\frac{\hat r}{\sigma \xi} \frac{1}{\sqrt{\pi}}\left( 1-\frac{2}{y_{21}^2} e^{-\frac{y_{21}^2}{4}}\right) \quad , \quad \quad \quad \; y_{21}/\sigma \gg 1 \;.
 \end{cases}
\ee 
Hence the variance grows linearly with the width for a small annulus,
and saturates for a large annulus to the value obtained above for the disk. This corresponds 
to a crossover of the variance from a volume to area law.

\vspace*{0.5cm}
\noindent{(iii) {\bf Gaussian shape linear statistics}.
One can also obtain explicit formula for $\phi(y)= e^{- \frac{a}{2}  y^2} $. In this case, it is possible to
compute explicitly the cumulants of arbitrary order from~\eqref{gen_cumul} using the formula
\be
\overline{ e^{- \frac{a}{2}  (y+v)^2}  } = f_a(y) = \frac{e^{- \frac{a y^2}{2 (1+ a \sigma^2)}}}{\sqrt{1 + a \sigma^2}}
\quad , \quad \int_{-\infty}^{+\infty} dy f_a(y)^n = \sqrt{\frac{2 \pi}{a n}} (1 + a \sigma^2)^{\frac{1-n}{2} } \;.
\ee 
Let us illustrate it on the variance. Using  
\bea 
&& \int_{-\infty}^{y_{\max}} dy \, \left( \overline{ \phi(y + v)^2}    - \left(  \overline{ \phi(y + v) } \right)^2  \right)  = \int_{-\infty}^{y_{\max}} dy \, \left( f_{2 a}(y)    - f_a(y)^2  \right) 
\eea 
one finds, for $\hat r<R$, 
\be
\langle {\cal L}^2_N \rangle \simeq 
 \sqrt{N} \frac{\hat r}{2 \sigma^2 \xi} \sqrt{\frac{2 \pi}{a }} (1 - \frac{1}{\sqrt{1 + a \sigma^2}} ) 
 = \sqrt{N}  \sqrt{\frac{2 \pi^2 \hat r^2 \rho_{\rm eq}(\hat r)}{a \sigma^2}} \left(1 - \frac{1}{\sqrt{1 + a \sigma^2}} \right)
\ee
Interestingly it vanishes at $a \sigma^2 \to 0$, and $a \sigma^2 \to +\infty$ and has
a maximum for $a \sigma^2= \frac{1}{2} (1 + \sqrt{5})$, i.e., the golden mean.





\subsection{Matching between macroscopic and microscopic linear statistics} \label{subsec:matching}

As we explained in the introduction, one of the motivation for this work is to understand 
how one can match the dependence $\sim N^{2-q}$ of the cumulants for the smooth linear statistics,
with the $\sim \sqrt{N}$ dependence of the cumulants for the counting statistics. Additionally
we want to have a picture of the fluctuations at the edge of the droplet which interpolates
between macroscopic and microscopic scale.

The first approach to these questions is to ask what happens if one inserts 
$f(r) = \phi( (r-\hat r)\sqrt{N}/\xi) $ into the formula for the cumulants obtained
for the smooth linear statistics recalled in Section \ref{subsec:previous}. Let us start with the first cumulant given in \eqref{first_cumul}.
If we inject in that formula $f(r)=\phi((r-\hat r) \sqrt{N}/\xi)$ we obtain (for $\hat r<R$)
\be \label{firstcum}
   \langle {\cal L}_N \rangle \simeq \sqrt{N} 
   2 \pi \hat r \, \rho_{\rm eq}(\hat r) \xi \int_{-\infty}^{y_{\rm max}}  dy \, \phi(y)  \quad, \quad y_{\max} = \begin{cases}
&+ \infty \;, \; \hat r < R \\
&0 \;, \; \quad \;\hat r = R \;.
\end{cases}  
\ee 
We can compare this expression with the result in \eqref{firstcumphi}, derived from the microscopic scales.
In the case $r< \hat R$, i.e. $y_{\max}=+\infty$, the latter reads
\begin{align}  \label{firstcumphi2}
&& \langle {\cal L}_N \rangle \simeq 
 \sqrt{N} \frac{\hat r}{2 \sigma^2 \xi}  \int_{-\infty}^{+\infty} dy \, \overline{\phi(y + v)} 
 \simeq \sqrt{N} \frac{\hat r}{2 \sigma^2 \xi}  \int_{-\infty}^{+\infty} dy \, \phi(y) 
\end{align}   
since one can simply shift the argument in the integral. 
Using the relation \eqref{xidef}, we see that it coincides exactly with \eqref{firstcum}.

In the case $\hat r= R$, i.e. $y_{\max}=0$, we find that there is an additional contribution from the edge
which is subdominant at small $\sigma$ (i.e. large $\xi$). Assuming that the function $\phi(y)$ is infinitely differentiable, one can expand 
\be \label{expansion} 
\phi(y + v) = \phi(y) + v \phi'(y) + \frac{v^2}{2}  \phi''(y) + \dots 
\ee 
and perform the Gaussian average as in Eq. \eqref{Gauss} using the result 
\be 
\frac{\overline{ v^{2 n}}}{(2 n)!} = \frac{\sigma^{2 n}}{2^n n!} \;.
\ee 
One then obtains the correction as a series in $\sigma$ in
terms of the derivatives of $\phi(y)$ at $y=0$ 
\begin{align}  \label{firstcumphi2}
\langle {\cal L}_N \rangle &\simeq 
 \sqrt{N} \frac{\hat r}{2 \sigma^2 \xi}  \int_{-\infty}^{0} dy \, \overline{\phi(y + v)} \\
&\simeq \sqrt{N} \frac{\hat r}{2 \sigma^2 \xi} \left(  \int_{-\infty}^{0} dy \, \phi(y) + 
 \sum_{n \geq 1 } \frac{\sigma^{2n}}{2^n n!}  \phi^{(2n-1)}(0) \right) \;.
\end{align}  
As discussed above, all these formulae are valid only when the test function $\phi$ is integrable. 

Let us now discuss the variance. Inserting $f(r)=\phi((r-\hat r) \sqrt{N}/\xi)$ in \eqref{secondcum2} we find
\begin{eqnarray} \label{secondcum3}
\langle {\cal L}_N^2 \rangle_c \simeq  \frac{\hat r}{2 \xi}  \sqrt{N} \int_{-\infty}^{y_{\max}} dy \, \phi'(y)^2  \;.
\end{eqnarray} 
Recall now the formula for the variance from the microscopic scales [see Eq. (\ref{second_cum})], 
\bea  \label{second_cum3}
&& \langle {\cal L}^2_N \rangle_c \simeq 
 \sqrt{N} \frac{\hat r}{2 \sigma^2 \xi}   \int_{-\infty}^{y_{\max}} dy \, \left( \overline{ \phi(y + v)^2}    - \left(  \overline{ \phi(y + v) } \right)^2  \right) \;.
\eea 
Its expansion to leading order in $\sigma$, using $\eqref{expansion}$ gives simply (since the $\sigma^2$ factors cancel to lowest order) 
\bea  \label{second4} 
&& \langle {\cal L}^2_N \rangle_c \simeq \sqrt{N} \frac{\hat r}{2  \xi} 
   \int_{-\infty}^{y_{\max}} dy \phi'(y)^2 \, (1 + O(\sigma^2) ) \;,
\eea  
which coincides to leading order in $\sigma$ with the formula \eqref{secondcum3}.

We can also obtain the corrections in powers of $\sigma$. In the case $r< \hat R$, i.e. $y_{\max}=+\infty$,
assuming that $\phi(y) \phi'(y)$ vanishes at infinity, 
one can shift one of the average and write
 \bea  \label{second_cum3}
\langle {\cal L}^2_N \rangle_c &\simeq& 
 \sqrt{N} \frac{\hat r}{2 \sigma^2 \xi}   \int_{-\infty}^{+\infty} dy \, \phi(y) (\phi(y) - \overline{\phi(y + v_2- v_1)} ) \\
&=& - \sqrt{N} \frac{\hat r}{2  \xi} 
\sum_{n \geq 1} \frac{ \sigma^{2 (n-1)}}{n!}  
\int_{-\infty}^{+\infty} dy \phi(y) \phi^{(2n)}(y) \;,
\eea 
where the leading order term reproduces \eqref{second4} upon integration by parts.
In the case $\hat r= R$, i.e. $y_{\max}=0$, one can express the subdominant
terms as a double series involving an integral of products of derivatives of $\phi(y)$.

Consider now the third cumulant. 
The formula from macroscopic scales
is given in \eqref{thirdcum2}. It involves only the 
derivative $f'(R)$ exactly at the edge of the droplet.
Hence inserting $f(r)=\phi((r-\hat r) \sqrt{N}/\xi)$
this formula gives 
\be \label{third_cumul}
\langle {\cal L}_N^3 \rangle_c \simeq \sqrt{N}  \frac{\sigma^2}{2} 
 \frac{R}{\xi}  \, \times \, \begin{cases} \phi'(+\infty)^3    \quad , \quad \hat r < R \\
 \phi'(0)^3   \quad , \quad \hspace*{1.5cm} \hat r = R 
 \end{cases} \;.
\ee 
In the case $\hat r < R$ and if $\phi'( {+\infty})=0$, this predicts that there is no sizeable contribution to the third cumulant from the bulk, 
i.e. $\langle {\cal L}_N^3 \rangle_c \simeq o(\sqrt{N})$, as confirmed below from a microscopic calculation. 
However, in the case $\hat r=R$,
one has $f'(R)=  \frac{1}{\xi } \sqrt{N} \phi'(0)$, hence it gives a non zero result.

On the other hand from \eqref{formula_cumul} we obtain formula for the third cumulant from the microscopic scales
\bea  \label{3cum3}
&& \langle {\cal L}^3_N \rangle_c \simeq 
\sqrt{N} \frac{\hat r}{2 \sigma^2 \xi}  \int_{-\infty}^{y_{\rm max}} dy  \overline {\left( \phi(y + v)  - \overline{\phi(y + v)} \right)^3 }
\eea  
Let us now insert the following expansion in powers of $\sigma$ (i.e., at small noise $v$)
\be 
\phi(y + v) -   \overline{\phi(y + v) }  =  v \phi'(y) + \frac{1}{2} ( v^2 - \sigma^2) \phi''(y)  + \frac{1}{6} v^3 \phi'''(y) + \dots
\ee 
This leads to
\bea  \label{exp} 
&& \overline{ \left( \phi(y + v) -   \overline{\phi(y + v) }  \right)^3  } 
= 3 \sigma^4  \phi'(y)^2 \phi''(y) \\
&& 
+ \sigma^6 \left( \phi''(y)^3  + 6 \phi'(y) \phi''(y) \phi'''(y) 
+ \frac{3}{2} \phi'(y)^2 \phi''''(y) \right) + 
O(\sigma^8) \;.
\eea 
By inserting this expansion in Eq. (\ref{3cum3}), one obtains
\bea \label{thirdcum}
\langle {\cal L}^3_N \rangle_c &\simeq& 
\sqrt{N} \frac{\hat r}{2  \xi} 3 \sigma^2 \int_{-\infty}^{y_{\rm max}} dy \phi'(y)^2 \phi''(y) \,   + O(\sigma^5) \nonumber \\
&=& \sqrt{N} \frac{\hat r}{2  \xi}  \sigma^2  (\phi'(y_{\max})^3 - \phi'(-\infty)^3)   + O(\sigma^5) 
\eea 
Hence we see that the leading term in the expansion in $\sigma$ is a total derivative ! 
Futhermore, in the case on which we focus here where $\phi'(-\infty)=0$, 
this leading term (which is $O(\sigma^3)$ since $\xi = O(1/\sigma)$) is exactly identical to the one in \eqref{third_cumul}, both for $\hat r < R$ (i.e., $y_{\max} = +\infty$) and for $\hat r = R$ (i.e., $y_{\max} = 0$).
This shows a perfect matching again in this case. For the special linear function $\phi(u) \propto u$
in which case $\phi'(-\infty) \neq 0$ we see from the microscopic formula \eqref{3cum3} and \eqref{thirdcum} 
that the third cumulant vanishes to $O(\sqrt{N})$. The corresponding macroscopic
calculation is performed in the \ref{app:linear}. 
\\

In \ref{app_cumul} we have obtained a formula for the general $q$-th cumulant to leading order in $\sigma$, from the
microscopic calculation, given in Eq. \eqref{CumqMicro}.
There, we also prove that this formula agrees with the general macroscopic formula for the $q$-th cumulant 
when inserting $f(r) = \phi( (r-R) \sqrt{N}/\xi)$. This shows the matching between microscopic and macroscopic
scales for arbitrary cumulants. 

\vspace*{0.5cm}
\noindent {\bf Remark 7}. We have shown that the matching 
between the microscopic formula and the macroscopic formula for the cumulants 
works to leading order for small $\sigma$ for the present case, i.e. for complex normal matrices $\beta=2$. 
For the more general Coulomb gas, there is no available microscopic formula at present, but in 
\cite{BLMS23} we have obtained the macroscopic formula for any $\beta >0$  and any space dimension. It is thus natural to {\it conjecture} that the corresponding matching property holds in
the general case, i.e. that inserting the form \eqref{def_micro} for $f(r)$  
yields the correct small $\sigma$ limit of the microscopic regime.




\section{Full probability distribution of the linear statistics} \label{sec:ldf}

In this section we study the PDF, ${\cal P} ( {\cal L}_N  )$, of the linear statistics ${\cal L}_N$ in the microscopic regime.
From the previous analysis of the cumulants, we expect that it takes the large deviation form in the large $N$ limit

\be \label{proba} 
{\cal P} ( {\cal L}_N  ) \sim \exp \left(  - \sqrt{N} \Psi\left(  \frac{  {\cal L}_N  }{\sqrt{N} } \right)   \right) \;,
\ee 
where $\Psi(w)$ is a rate function which we now determine. 

\subsection{General expression of the rate function $\Psi(w)$}

We restrict oursemves here to the case where the test function
$\phi(u)$ decays sufficiently fast, specifically we require $\lim_{u \to \pm \infty} u^{1+ \eta}\, \phi(u) = 0$ for some $\eta>0$. 
In that 
case the first moment \eqref{firstcum} is finite (i.e.,  it scales as $\sqrt{N}$) and we write, in the large $N$ limit 
\be 
\tilde \chi(\mu,N) = \sqrt{N} \tilde \chi(\mu) + o(\sqrt{N}) \;,
\ee 
where, according to the result obtained previously in \eqref{formula_cumul}, one has
\be \label{chi_tildemu2}
\tilde \chi(\mu)= 
- \frac{\hat r}{2 \xi \sigma^2} \int_{-\infty}^{y_{\rm max}}  dy \log \overline{ e^{- \mu \phi(y+v)}} \;,
\ee 
recalling that the overbar denotes a Gaussian average over $v$ [see \eqref{Gauss}] and $y_{\rm max}=+\infty$ for $\hat r<R$ (in the bulk) while $y_{\rm max}=0$ for $\hat r=R$ (at the edge).
Inserting the form \eqref{proba}
into \eqref{eq:chidef} 
we obtain 
\be 
\tilde \chi(\mu) = - \min_{w \in \mathbb{R}} (\mu w + \Psi(w))  \;.
\ee 
Upon Legendre inversion, we obtain the rate function associated to the large deviation form in (\ref{proba}) as 
\be \label{Psi} 
\Psi(w) = - \min_{\mu \in \mathbb{R}} (\mu w + \tilde \chi(\mu) )  \;.
\ee 
We can thus obtain $\Psi'(w)$ in a parametric form as
\begin{align} \label{param} 
\Psi'(w) = - \mu  \quad , \quad  
w = - \tilde \chi'(\mu) = - \int_{-\infty}^{y_{\rm max}}  dy \frac{\overline{ \phi(y+v) e^{- \mu \phi(y+v)}} }{\overline{ e^{- \mu \phi(y+v)}}} \;,
\end{align} 
from which one can in principle obtain the asymptotic behaviors of $\Psi(w)$ for small and large $w$. We now turn our focus to this analysis.

\subsection{Asymptotic behavior of the rate function}

We will not study the function $\Psi(w)$ in details but only give its asymptotic behaviors. The function $\Psi(w)$ admits a minimum around the (scaled) average of ${\cal L}_N$, namely near $w = \Delta_1$, where $\Delta_1 = \lim_{N \to +\infty} \langle {\cal L}_N\rangle/\sqrt{N}$ -- see Eq. (\ref{firstcumphi}). Near this minimum, $\Psi(w)$ takes a quadratic form, namely
\be \label{psi_gauss}
\Psi(w) \underset{w \to \Delta_1}\simeq  \frac{1}{2 \Delta_2} (w- \Delta_1)^2 
\quad , \quad \Delta_q = \lim_{N \to +\infty} \frac{\langle {\cal L}_N^q \rangle^c}{\sqrt{N}} \;,
\ee 
where $\Delta_2$ can be read off from \eqref{second_cum}. 

We now discuss the behavior of $\Psi(w)$ for $w \to + \infty$ and $w \to - \infty$. These limits 
correspond respectively to $\mu<0$ and to $\mu>0$. The cases $\mu \to - \infty$ and to $\mu \to + \infty$
are controled respectively by
the behavior of the maximum (and minimum) value of $\phi$. Let us start with the limit $w \to + \infty$. 
We assume that $\phi_{\max} = \max_{u \in \mathbb{R}} \phi(u) >0$ and is reached only within a
region of finite size $u=O(1)$. From \eqref{param} we see that we must study the behavior of $\tilde \chi(\mu)$ for $\mu \to -\infty$.
We start from Eq. (\ref{chi_tilde_bulk}), which reads, with $\mu = - |\mu|$
\begin{align}
&\tilde \chi(\mu,N) \simeq \sqrt{N} \frac{\hat r}{2 \xi} \int_{-\infty}^{\tilde \lambda_{\max}} \ln \frac{J(\mu, \tilde \lambda)}{J(0,\tilde \lambda)} d\tilde \lambda 
\quad, \quad \label{chi_tilde_ratio}\\
&\frac{J(\mu, \tilde \lambda)}{J(0,\tilde \lambda)} = e^{-\tilde \lambda^2 \frac{\sigma^2}{2}}\frac{1}{\sqrt{2 \pi \sigma^2}} \int_{-\infty}^\infty du e^{- \frac{u^2}{2 \sigma^2} + \tilde \lambda u + |\mu| \phi(u)} \label{ratio_J}
\end{align}
where we recall that $\tilde \lambda_{\max}=+\infty$ for $\hat r<R$ (bulk) and $\tilde \lambda_{\max}=0$ for $\hat r=R$ (edge).
We show in the following that the ratio in (\ref{ratio_J}) takes a large deviation form, for $|\mu| \to \infty$
\begin{align} \label{large_dev}
\frac{J(\mu, \tilde \lambda)}{J(0,\tilde \lambda)} \simeq e^{|\mu| \varphi\left( \zeta = \frac{\tilde \lambda}{\sqrt{\mu}}\right)} \quad, \quad \varphi(\zeta) = 
\begin{cases}
&\phi_{\rm max} - \frac{\sigma^2}{2}\,\zeta^2 \quad, \quad |\zeta| \leq \zeta_c = \sqrt{\frac{2 \phi_{\rm max}}{\sigma^2}} \\
&0 \quad, \quad \hspace*{2.cm} |\zeta| \geq \zeta_c  
\end{cases} \quad \;.
\end{align}
\\
Setting $\tilde \lambda = \zeta \sqrt{|\mu|}$, the ratio in (\ref{ratio_J}) reads
\begin{align} \label{calJ}
\frac{J(\mu, \tilde \lambda = \zeta \sqrt{|\mu|})}{J(0,\tilde \lambda = \zeta \sqrt{|\mu|})} = e^{-|\mu| \zeta^2 \frac{\sigma^2}{2}} {\cal J}(\mu,\zeta) \quad, \quad {\cal J}(\mu,\zeta) = \int_{-\infty}^{\infty} \frac{du}{\sqrt{2 \pi \sigma^2}} e^{- \frac{u^2}{2 \sigma^2} + \zeta \sqrt{|\mu|} u + |\mu| \phi(u)} \;.
\end{align}
In the large $|\mu|$ limit, it turns out that there are two distinct regions that contribute to the integral over $u$ in (\ref{calJ}), both contributions being of order $O(e^{|\mu|})$: 
\begin{enumerate}
\item[(1)]{Region 1: for $u=O(1)$ in the region where the maximum of $\phi(u)$ is reached,  
the argument of the exponential in (\ref{calJ}) is dominated by the third term. This leads to a first contribution (assuming $\zeta \ll \sqrt{|\mu|}$)
\bea \label{calJ_1}
{\cal J}(\mu,\zeta)  \vert_1 = e^{|\mu| \phi_{\rm max} + o(|\mu|)} \;, \; \mu \to -\infty \;.
\eea
}
\item[(2)]{Region 2: for large $u = O(\sqrt{|\mu|})$ where the argument of the exponential in (\ref{calJ}) is instead dominated by the two first terms, which are both of the same order, namely $O(|\mu|)$, while the third term is subleading -- because of our assumptions on the large $u$ asymptotic behavior of $\phi(u)$.
This leads to the second contribution
\bea \label{calJ_2}
{\cal J}(\mu,\ell)  \vert_2 = e^{|\mu| \zeta^2 \frac{\sigma^2}{2} + o(|\mu|)} \;, \; \mu \to -\infty \;.
\eea
}
\end{enumerate}
Therefore, summing these two contributions (\ref{calJ_1}) and (\ref{calJ_2}), one obtains
\be \label{calJ_sum}
{\cal J}(\mu,\zeta) = {\cal J}(\mu,\zeta)  \vert_1 + {\cal J}(\mu,\zeta)  \vert_2  = e^{|\mu|\,\max\left(\phi_{\max}, \zeta^2 \frac{\sigma^2}{2}\right) + o(|\mu|)} \;, \; \mu \to -\infty \;.
\ee
Finally, by injecting this asymptotic behavior (\ref{calJ_sum}) in Eq. (\ref{calJ}), one obtains the result given in Eq. (\ref{large_dev}).

To proceed, we inject this asymptotic behavior (\ref{large_dev}) in the formula for $\tilde \chi(\mu,N) \simeq \sqrt{N} \tilde \chi(\mu)$ in Eq. (\ref{chi_tilde_bulk}) and one finds, 
using Eq. (\ref{xidef}), as $\mu \to -\infty$
\begin{align} \label{chi_mu_large_bulk2}
\tilde \chi(\mu) \simeq B_-  |\mu|^{3/2} \quad , \quad B_-=  \frac{4}{3} [\phi_{\rm max}]^{3/2} \, \hat r \, \sqrt{2 \pi \rho_{\rm eq}(\hat r) } 
\times 
\begin{cases} 
1 \quad, \quad \; \hat r < R \\
\frac{1}{2} \quad, \quad \hat r = R 
\end{cases}
\;.
\end{align}
Remarkably the prefactor is independent of $\sigma$. It is interesting to compare this result
with the one obtained in \cite{castillo} in the case of the counting statistics $\phi(u)=\theta(-u)$ (within the bulk).
This formula (\ref{chi_mu_large_bulk2}) is in agreement with Eqs. (30) and (34) of Ref. \cite{castillo}, up to a factor of $2$. This is not surprising since our assumptions on $\phi(u)$ exclude this
particular choice where the first moment does not scale as $\sqrt{N}$. It is easy to extend our calculation
to also cover that case, but we will not detail it here.

Concerning the limit $w \to - \infty$, there are two cases to consider: (i) $\phi_{\min} = \min_{u \in \mathbb{R}} \phi(u) <0$ 
and (ii) $\phi_{\min} = \min_{u \in \mathbb{R}} \phi(u) =0$. We consider for now only the first case (i). 
In that case the same result as in \eqref{chi_mu_large_bulk2} holds with the replacement $B_- \to B_+$ and $\phi_{\rm max} \to |\phi_{\rm min}|$. As in Ref. \cite{castillo}, the tail $\tilde \chi(\mu)| \propto |\mu|^{3/2}$ immediately leads to a cubic behavior of the rate function for $w \to \pm \infty$ from \eqref{param} and one obtains 
\be \label{cubic}
\Psi(w) \simeq A_\pm |w|^3 \quad , \quad A_\pm = \frac{4}{27 B_\mp^2} \;,
\ee 
where the amplitude $A_+$ is thus given by 
\bea \label{Apm}
A_+=  \frac{1}{24 \pi} \frac{1}{\phi_{\max}^3 \hat r^2 \rho_{\rm eq}(\hat r)}
\times 
\begin{cases} 
1 \quad, \quad \; \hat r < R \\
4 \quad, \,\quad \hat r = R
\end{cases} \quad, 
\eea
while $A_-$ is given by the same expression with the substitution $\phi_{\max} \to \phi_{\min}$. Finally in the case $\min_{u \in \mathbb{R}} \phi(u) =0$ since ${\cal L}_N >0$ the PDF ${\cal P}({\cal L}_N)$ has a support only
on the positive real axis, hence the rate function is formally infinite for $w>0$. This is consistent
with the fact that the amplitude $A_+$ diverges as $\sim |\phi_{\rm min}|^{-3}$ as 
$|\phi_{\rm min}|\to 0$.

\vspace*{0.5cm}
\noindent 
{\bf Remark 8. Scaling form:} One can understand why the leading asymptotics of 
$\tilde \chi(\mu,N)$ at large $\mu \to - \infty$ in Eq. (\ref{chi_mu_large_bulk2}) does not depend on $\sigma$. 
Writing the CGF $\tilde \chi(\mu,N) \simeq \sqrt{N} \tilde \chi(\mu)$ in the scaling form
\bea \label{scaling}
\tilde \chi(\mu) = \frac{1}{\sigma^3} {\cal F}(\hat \mu = \mu \sigma^2, \sigma) \;,
\eea
one can show that in 
the limit $\sigma \to 0$ with $\hat \mu = \mu \sigma^2$ fixed, it becomes
\bea \label{scaling2}
\tilde \chi(\mu) \simeq \frac{1}{\sigma^3} {\cal F}_0(\hat \mu = \mu \sigma^2) \;,
\eea
which reproduces the $\sigma$-dependence of the cumulants in the limit $\sigma \to 0$ -- see Eq. (\ref{CumqMicro}). 
Note also that if one assumes that ${\cal F}_0(z) \sim |z|^{3/2}$ as $z \to - \infty$, one finds that this scaling form (\ref{scaling2}) implies $\tilde \chi(\mu) \propto |\mu|^{3/2}$, independently of $\sigma$, which is in agreement
with~(\ref{chi_mu_large_bulk2}).


\vspace*{0.5cm}
\noindent
{\bf Remark 9}. It is interesting to 
perform the asymptotic analysis for $\mu \to - \infty$, which leads to \eqref{chi_mu_large_bulk2},
starting instead from the formula involving the constrained density \eqref{dchitilde_txt}.
The detailed calculation of the constrained density is performed in \ref{App:const_density}.
One finds that for $\sigma>0$ the constrained density develops a peak near the maximum of $\phi(u)$, of 
width $u \sim 1/\sqrt{|\hat \mu|}$, and height $\propto |\hat \mu|$ where $\hat \mu= \mu \sigma^2$. 
This peak is 
accompanied
by a depletion region of size $\sqrt{|\hat \mu|}$ around the peak, see Fig. \ref{fig_density}. In the double limit $\mu \to - \infty$, $\sigma \to 0$ keeping $\hat \mu$ fixed, the density is approximated by
\be \label{densityhatmu}
\rho_\mu(u) \simeq \frac{1}{\pi} \left(1 - |\hat \mu| \phi''(u) \right) \;,
\ee 
for $|\hat \mu| < \mu_c$ such that Eq. \eqref{densityhatmu} is positive everywhere. 
For $|\hat \mu|>\mu_c $ a hole develops where the density vanishes
and the support of the density becomes a disk plus an annulus (for more details see \ref{App:const_density}).

\subsection{Matching with the Coulomb gas}

In the previous section we found that the rate function $\Psi(w)$ 
in \eqref{proba} associated to the PDF in the microscopic regime (for ${\cal L}_N = O(\sqrt{N})$), exhibits a cubic behavior $\sim w^3$ for $w \to + \infty$ [see Eq. (\ref{cubic})]. 
This cubic behavior is similar to the one found for the counting statistics in Ref. \cite{castillo} (up to an amplitude as discussed in the previous section).
In that case it was shown that this cubic behavior matches the small argument behavior
of the rate function for the PDF in the macroscopic Coulomb gas (CG) regime (for ${\cal L}_N = O(N)$), i.e. 
More precisely in the CG regime one expects the large deviation form
\be \label{probaCG} 
{\cal P} ( {\cal L}_N  ) \sim \exp \left(  - N^2 \Psi_{\rm CG}\left(   \Lambda = \frac{  {\cal L}_N  }{N } \right)   \right) \;.
\ee 
Note that in previous papers on the counting statistics \cite{allez,castillo} the scaling variable $\Lambda$ was denoted $\kappa$.
The matching between the microscopic and macroscopic regimes requires that $\Psi_{\rm CG}(\kappa) \sim \Lambda^3$ as $\Lambda \to 0$, which was checked explicitly in Refs. \cite{Lacroix_rotating,castillo}. This cubic behavior at small argument may seem surprising at first sight, since in the CG approach, whenever the constrained density is smooth, one obtains instead a quadratic behavior 
\cite{BLMS23}. This however holds only for sufficiently smooth linear statistics, while 
for the counting statistics it was shown that the constrained density develops a singular behavior (namely a delta peak next to a hole where the density vanishes), leading instead to a cubic behavior \cite{allez}.

We now show that a similar phenomenon is at play in the present case for a larger class of linear statistics
with a function $f(r)$ of the form \eqref{def_micro} which varies at the microscopic scale. 
As anticipated from the last remark in the previous section, we find that 
this microscopic perturbation results in a {\it macroscopic} hole in the constrained density,
in the matching regime $\Lambda \to 0$. 

To show this we consider the constrained minimization problem of the 
Coulomb gas energy 
\bea \label{energy}
{\cal E}[\rho] = \int d^2 {\bf r} \rho({\bf r}) U({\bf r}) - \frac{1}{2} \int d^2 {\bf r} \int d^2 {\bf r'} \rho({\bf r}) \rho({\bf r'}) \ln(|{\bf r}-{\bf r'}|) \;,
\eea
under the constraints
\begin{align} 
&\int d^2 {\bf r} \rho({\bf r}) = 1 \;,\label{norm_CG} \\
&\int d^2 {\bf r} \rho({\bf r}) f({\bf r}) = \int d^2 {\bf r} \rho({\bf r}) \phi\left(\sqrt{N}\frac{(r-\hat r)}{\xi}\right) = \Lambda \;. \label{const}
\end{align}
We first determine the optimal density $\rho_\Lambda$, and from it we obtain 
\be \label{Psi_CG}
\Psi_{\rm CG}(\Lambda) = \beta ( {\cal E}[\rho_\Lambda] - {\cal E}[\rho_{\Lambda = 0}] ) \;,
\ee 
where we substracted the energy of the ground state for $\rho_{\Lambda= 0}=\rho_{\rm eq}$. 
Here we restrict to the Ginibre case, i.e. $\beta=2$ and $U(r)=r^2/2$, although it
is straightforward to extend our calculation to the general case (for arbitrary $U(r)$ and general $\beta>0$).

\begin{figure}[t]
\centering
\includegraphics[width = 0.6 \linewidth]{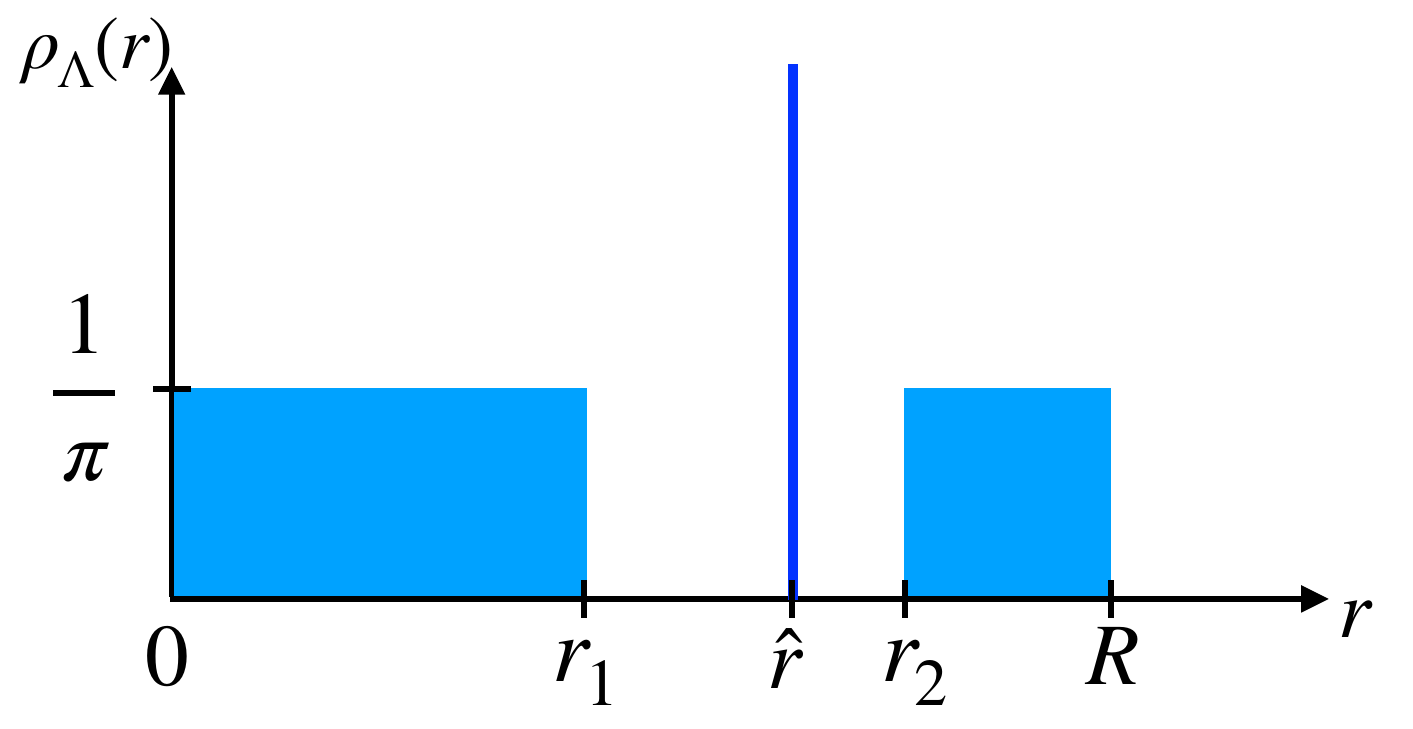}
\caption{Plot of the conditioned density $\rho_\Lambda({\bf r}) \equiv \rho_\Lambda(r=|{\bf r}|)$ given in Eqs. (\ref{rho_ansatz}) and (\ref{def_I}). The vertical blue line indicates the delta function at $r=\hat{r}$.}\label{Fig_hole}
\end{figure}

Based on the microscopic result (see Fig. \ref{fig_density}), and the analysis for the counting statistics \cite{allez}, it is natural to look for a solution $\rho_\Lambda({\bf r})$ of the form (see Fig. \ref{Fig_hole})
\bea \label{rho_ansatz}
\rho_\Lambda({\bf r}) \equiv \rho_\Lambda(r = |{\bf r}|) = \frac{\Lambda}{2 \pi \hat r \phi_{\rm max}} \delta(r-\hat r) + \frac{1}{\pi} I_{r_1,r_2}(r)
\eea
where 
\bea \label{def_I}
I_{r_1, r_2}(r)=
\begin{cases}
&1 \;, \; 0<r<r_1 \\
& 0 \;, \; r_1 < r < r_2 \\
& 1 \;, \; r_2 < r < R \;.
\end{cases}
\eea
The delta function part corresponds to the peak in Fig. \ref{fig_density} (for $\hat \mu \to -\infty$). 
Its amplitude is dictated by the constraint \eqref{const} since the integral is over a microscopic scale
of order $1/\sqrt{N}$. The value of the density outside the hole is dictated by the local equilibrium
in the absence of the constraint (it gives a negligible contribution to the integral \eqref{const}).
The only variational parameters are thus the radius of the disk $r_1$, together
with the radii of the annulus $r_2$ and $R$ (which are related through the normalization
condition~\eqref{norm_CG}). 

The minimization of the energy ${\cal E}[\rho]$ with respect to $r_1,r_2$ and $R$ is performed
in \ref{sec:CG}. 
The rate function $\Psi_{\rm CG}(\Lambda)$ can then be computed from Eq. (\ref{Psi_CG}) and 
its exact form is displayed in \ref{sec:CG}. Here we give its 
small $\Lambda$ behavior~(see~\ref{sec:CG}) 
\bea 
\Psi_{\rm CG}(\Lambda) \simeq \frac{1}{24} \frac{\Lambda^3}{\phi_{\rm max}^3 \hat r^2} + \dots  \quad, \quad \Lambda \to 0  \;. \label{small_lamb}
\eea 
We see that both the power law behavior and the prefactor match exactly the $w \to +\infty$ asymptotics 
of $\Psi(w)$ 
obtained in
\eqref{cubic}-\eqref{Apm}, i.e, one has in the matching region
$ N^2 \Psi_{\rm CG}(\Lambda) \simeq \sqrt{N} \Psi(w=\sqrt{N} \Lambda)$. 

For finite $\Lambda$, the solution of the minimization problem has a richer structure. In all cases we find that $R=1$ at the optimum.
For $\phi(u)>0$ we find that the rate function is well defined in a finite region in the plane $(\hat r, \tilde \Lambda = \Lambda/\phi_{\max}) \in [0,1] \times [0, 1]$ and smooth in a subregion containing $\Lambda=0$ and 
bounded by two curves where the optimal $r_1$ and $r_2$ reach $r_1=0$ and $r_2=1$ respectively (see Fig. \ref{Fig_PhDiag}). 
In total there are three distinct phases. 
The explicit formulae for $\Psi_{\rm CG}(\Lambda)$ in each phase are given in \ref{sec:CG}, together
with the critical behavior across the transition lines. 

\section{Conclusion} 

In this paper we studied the Coulomb gas in 2D for $\beta=2$ in a confining (rotationally invariant) potential $N U(r)$,
which describes the eigenvalues of normal complex random matrices (including the complex Ginibre ensemble, that corresponds to $U(r)=r^2/2$). In this case the equilibrium density, for large $N$, is supported on the disk centered at the origin, of radius $R = O(1)$. 
We studied the microscopic linear statistics associated to a test function of the form $f(r) = \phi(\sqrt{N} (r - \hat r)/\xi)$,
where $\phi(u)$
and $\xi$ are fixed in the limit $N \to +\infty$. We obtained a general formula for the
variance and for the higher cumulants in that limit, which we find to be of order $\sqrt{N}$,
with a coefficient given explicitly by some nontrivial integrals involving the function $\phi(u)$.
We showed that when $\xi \to +\infty$ this formula matches precisely
at leading order the corresponding results obtained in \cite{BLMS23} for smooth linear statistics.
Our result demonstrates that the leading contribution to the cumulants of order $q \geq 3$ in this limit comes only from a boundary layer of size $1/\sqrt{N}$ near the edge of the droplet. We also showed that there are two distinct regimes 
for the large deviation forms of the full distribution of ${\cal L}_N$ which we computed explicitly: (i) for ${\cal L}_N = {O}(\sqrt{N})$ and (ii) for ${\cal L}_N = O(N)$. As in the case of the FCS, the second regime
corresponds to a singular optimal distribution of charges, with a delta peak in the density surrounded by
holes. Our approach has allowed to describe in detail how this distribution
of charges arises from the microscopic scales.

There are possible applications and extensions to the present work. An immediate one is the case of
the Ginibre symplectic ensemble (GinSE) for which there is an analogous exact formula for 
the CGF \cite{Rider}, see Eq. (61) in \cite{BLMS23}. Comparing with \eqref{det_form} in the present work 
we see that the replacement is simply $U(r)=r^2$ and $\ell \to 2 \ell$. Following
the same steps as here we obtain that for the microscopic linear statistics 
\eqref{def_micro}} with a given $\xi$, the CGF is given by $\tilde \chi_{\rm GinSE}(\mu,N)= \frac{1}{2}
\tilde \chi(\mu,N)|_{U(r)=r^2}$, from which the cumulants can be obtained. 

Another interesting application is to non-interacting fermions in two dimensions in a rotating harmonic potential $V(\tilde r) = \frac{1}{2} m \omega^2 \tilde r^2$, where
$\tilde r$ denotes the distance to the origin (and here $m$ is the mass of the fermions). When the rotating frequency $\Omega$
approaches the trap frequency $\omega$, i.e. $(\Omega/\omega) -1 = O(1/N)$, the 
ground state of the system maps onto the lowest Landau level of electrons in an external magnetic field. 
In that case the many body quantum probability of the positions of the fermions is exactly given by the joint distribution of the eigenvalues of the complex Ginibre ensemble, i.e., by Eq.~(\ref{PDF_intro}) with $U(r)=r^2/2$,
with the correspondence $\tilde r_i = r_i \sqrt{ N \hbar/m \omega}$~\cite{Lacroix_rotating}. The linear statistics
considered here amounts to study the fluctuations of the quantum average in the ground state of the microscopic observable 
${\cal L}_N = \sum_{i=1}^N \phi((\tilde r_i - \hat r)/\xi)$~\cite{Lacroix_rotating,Smith_rotating,Manas_rotating1,Manas_rotating2}. One can imagine
that such linear statistics could be measured by local probes such as the tip of a microscope,
or in cold atoms experiments, by direct imaging of the positions of the fermions using a quantum microscope~\cite{Cheuk:2015,Haller:2015,Parsons:2015,tarik2024}. 
Yet another interesting problem would be to study non-interacting fermions on the sphere, 
see \cite{Forrester} Chapter 15, especially in presence of point charges, which
were recently considered in the Coulomb gas context \cite{Byun2024,Byun2025}.

Finally, it would be interesting to extend our study to the case where the equilibrium
Coulomb gas has a multiple support, i.e. a collection of annuli centered at the origin. Indeed in 
that case it has been found that there are additional mesoscopic fluctuations of the charges
involving the different components of the support \cite{Charlier22,Charlier23}. 

\bigskip

{\bf Acknowledgments.} We thank P. Bourgade for useful discussions. We acknowledge support from ANR Grant No. ANR-23-CE30-0020-01 EDIPS. 

\appendix

\section{Exact solution for the linear case $f(r) \propto r$ and Ginibre ensemble}\label{app:linear}

For the linear statistics $f_N(r)=c \, r$, and for the Ginibre ensemble $U(r)=r^2/2$, one can perform all the integrals
exactly in Eq. (\ref{det_form}) and one obtains 
\be 
\tilde \chi(\mu,N) = \sum_{\ell=0}^{N-1} \log \frac{\int_0^{+\infty} dr r^{2 \ell+1} e^{- N r^2 - \mu c r}}{\int_0^{+\infty} dr r^{2 \ell+1} e^{- N r^2}}
= \sum_{\ell=0}^{N-1} \log \frac{ U( \ell+1 , 1/2, c^2 \mu^2/(4 N))  }{U( \ell+1 , 1/2, 0)  } \;,
\ee 
where $U(a,b,z)$ is the Tricomi confluent hypergeometric function and where we have used the identity (for $A>0$)
\be  \label{U} 
\int_0^{+\infty} dr r^A e^{-r^2 - \mu r} 
= \Gamma(1+A) 2^{-(1+A)} U\left( \frac{A+1}{2} , 1/2, \mu^2/4 \right) \;.
\ee 

Let us consider the regime $s= \mu/N = O(1)$ and $c=O(1)$. In that case the
sum over $\ell$ is dominated by large $\ell$, and one can set $\ell= \lambda N$ and then use the asymptotic behavior of the hypergeometric function $U(a,b,z)$ for large paramater $a$ given in the formula 13.8.8 of \cite{DLMF_U}. After a simple calculation this leads to 
\bea \label{chi(s)}
\chi(s)= \lim_{N \to +\infty} 
\frac{1}{N^2} \chi(s,N) = \frac{c^2 s^2}{8} - 2  \int_0^1 d \lambda  \beta(s,\lambda) \lambda  
\eea 
where 
\be 
2 \beta(s,\lambda) = w + \sinh w \quad {\rm with} \quad \cosh w = 1 + \frac{s^2 c^2}{8 \lambda} \;.
\ee 
We find after evaluating the integral in (\ref{chi(s)})
\be 
\chi''(s) = \frac{1}{32} c^2 \left(c s \left(c s-\sqrt{c^2 s^2+16}\right)+8\right) \;.
\ee 
This agrees with the result from \cite{BLMS23}  (for $c=1$ see Eq. (A.8) there upon
correcting a misprint in Eq. (A.9)), which reads in parametric form
\be 
\chi'' = \frac{R_s^2}{4} \quad , \quad R_s^2 + \frac{s}{2} R_s = 1 \;. 
\ee 
This leads to the first three cumulants
\be 
\langle {\cal L} \rangle = \frac{2 c}{3} N  \quad , \quad \langle {\cal L}^2 \rangle = \frac{c^2}{4} 
\quad , \quad \langle {\cal L}^3 \rangle = \frac{c^3}{8}  N^{-1} \;.
\ee

\section{Higher cumulants}\label{app_cumul}

Here we study the limit of small $\sigma$ of the formula \eqref{chi_tilde_bulk}. To this aim we 
perform the rescaling 
\be 
\mu = \frac{1}{\sigma^2} \hat \mu \quad , \quad \tilde \lambda = \frac{1}{\sigma^2} \hat \lambda \;,
\ee 
which leads to 
\be \label{def_K}
\tilde \chi(\mu,N) \simeq \sqrt{N} \frac{\hat r}{2 \xi \sigma^2} \int_{-\infty}^{\hat \lambda_{\rm max}} d\hat \lambda
\log \frac{  K(\hat \mu,\hat \lambda) }{K(0,\hat \lambda)}  
~,~ K(\hat \mu,\hat \lambda)=  \int_{-\infty}^{+\infty} du
e^{ - \frac{1}{\sigma^2} (  \frac{u^2}{2} -  \hat \lambda u   + \hat \mu \phi(u))  } \;,
\ee 
where
\be \label{def_lmax2}
\hat \lambda_{\rm max} = \begin{cases}  
 +\infty \quad \hat r < R \\
0 \quad \quad \,\hat r = R \;.
\end{cases}
\ee 

In the limit $\sigma \ll 1$ we can use the saddle point method to evaluate the integral over $u$ in (\ref{def_K}). It leads to the asymptotic estimate
\be \label{chi_small_sigma}
\tilde \chi(\mu,N) \simeq - \sqrt{N} \frac{\hat r}{2 \xi \sigma^4} \int_{-\infty}^{\hat \lambda_{\rm max}} d\hat \lambda
\, \left( \min_{u \in \mathbb{R}} \left[ \frac{u^2}{2} - \hat \lambda u + \hat \mu \phi(u) \right] - \min_{u \in \mathbb{R}} \left[ \frac{u^2}{2}-  \hat \lambda u   \right ] \right) \;.
\ee 
Hence, introducing the saddle point value for $u$
\be \hat \lambda = u + \hat \mu \phi'(u)  \quad \Longleftrightarrow \quad u=u_{\hat \mu,\hat \lambda}
\ee 
and taking a derivative w.r.t. $ \mu$ we obtain {\red we assume a single root} 
\be 
\partial_{ \mu} \tilde \chi(\mu,N) = \sigma^2 \partial_{\hat \mu} \tilde \chi(\mu,N)  \simeq - \sqrt{N} \frac{\hat r}{2 \xi \sigma^2} 
\int_{-\infty}^{\hat \lambda_{\rm max}} d\hat \lambda \, 
\phi(u_{\hat \mu,\hat \lambda}) \;.
\ee 
Performing similar manipulations as in Section \ref{subsec:previous} and in \cite{BLMS23} we obtain
\be
\partial_{ \mu} \tilde \chi(\mu,N) 
\simeq  - \sqrt{N} \frac{\hat r}{2 \xi \sigma^2} \int_{-\infty}^{\hat \lambda_{\rm max}} 
d ( u_{\hat \mu,\hat \lambda} +  \hat \mu \phi'(u_{\hat \mu,\hat \lambda})   )  \, 
\phi( u_{\hat \mu,\hat \lambda}) \;. \label{trick} 
\ee
Let us first consider the case $\hat r< R$, in which case $\hat \lambda_{\rm max}=+\infty$.
Performing the change of variable $y= u_{\hat \mu,\hat \lambda}$ we obtain, at large $N$ and to leading order in $\sigma$ (at fixed $\hat \mu$) 
\be 
\partial_{\mu} \tilde \chi(\mu,N) \simeq - \sqrt{N} \frac{\hat r}{2 \xi \sigma^2} \int_{-\infty}^{+\infty} dy \, \partial_y ( y + \hat \mu  \phi'(y)   )  \, 
\phi(y)   \;.
\ee
Using \eqref{formula_cumul} and recalling that $\hat \mu = \sigma^2 \mu$, we see that 
this result agrees with the formula given in \eqref{firstcumphi2} and \eqref{second4} 
for the first and second cumulant (up to an integration by part using that $\phi$
and $\phi'$ vanish at infinity). Furthermore, we see that
to this leading order in $\sigma$, there is no contribution from the bulk to the higher cumulants, i.e. for $\hat r < R$
(e.g. the term of order $O(\sigma^3)$ in the third cumulant vanishes).

Consider now the edge regime $\hat r= R$. In that case, after the change of variable $y= u_{\hat \mu,\hat \lambda}$ from \eqref{trick}
one obtains instead
\be 
\partial_{\mu} \tilde \chi(\mu,N) \simeq - \sqrt{N} \frac{R}{2 \xi \sigma^2} \int_{-\infty}^{u_{\hat \mu}} dy \, \partial_y ( y + \hat \mu  \phi'(y)   )  \, 
\phi(y)  \;, \label{a1} 
\ee
where the upper boundary of integration is 
$u_{\hat \mu}=u_{\hat \mu,\hat \lambda=0}$, which is the solution of 
\be \label{saddle_1}
u = - \hat \mu \phi'(u) \quad \Longleftrightarrow \quad u=u_{\hat \mu} \;.
\ee 
Setting $\mu=0$ in the above formula \eqref{a1}, and since $u_0=0$, we recover the formula \eqref{firstcumphi2} to leading order in $\sigma$. 
Let us now perform an integration by part from \eqref{a1} which leads~to
\be 
\partial_{\mu} \tilde \chi(\mu,N) \simeq  \sqrt{N} \frac{R}{2 \xi \sigma^2} \int_{-\infty}^{u_{\hat \mu}} dy \,  ( y + \hat \mu  \phi'(y)   ) \phi'(y)   \, \;,
  \label{a2}  
\ee 
since the boundary term at $y=u_\mu$ vanishes thanks to \eqref{saddle_1} and we assume that $y \phi(y)$ and $\phi'(y) \phi(y)$
vanish at $y=-\infty$. Taking another derivative w.r.t. $\mu$ leads to
\be 
\partial_{\mu}^2 \tilde \chi(\mu,N) \simeq  \sqrt{N} \frac{R}{2 \xi } \int_{-\infty}^{u_{\hat \mu}} dy \,   \phi'(y)^2    \,.
  \label{a3}  
\ee 
Setting $\mu=0$ we recover the formula \eqref{second4} to leading order in $\sigma$.

Taking a derivative of \eqref{saddle_1} one finds 
\be 
 \frac{d u_{\hat \mu} }{d \hat \mu}  = - \frac{\phi'(u_{\tilde \mu})}{ 1 + \tilde \mu \phi''(u_{\tilde \mu}) }
= \frac{ \phi'(u_{\tilde \mu})^2}{u_{\tilde \mu} \phi''(u_{\tilde \mu}) - \phi'(u_{\tilde \mu})} \;.
\ee 
Using $\partial_\mu = \sigma^2 \partial_{\hat \mu} = \sigma^2 \frac{d u_{\hat \mu} }{d \hat \mu} \partial_{\hat \mu}$ we
obtain 
\begin{align} \label{CumqMicro}
 \langle {\cal L}_N^q \rangle_c &= (-1)^{q}  
\partial_\mu^q \tilde \chi(\mu,N)|_{\mu=0} 
=  (-1)^{q}  \sigma^{2 (q-2)} \partial_{\hat \mu}^{q-3} \left( \partial_{\hat \mu} \partial_\mu^2\tilde \chi(\hat \mu,N)\right) |_{\hat \mu=0} \\
&\simeq (-1)^{q}  \sigma^{2(q-2)}  \sqrt{N} \frac{R}{2 \xi} 
(\frac{ \phi'(u)^2}{u \phi''(u) - \phi'(u)} \partial_u)^{q-3} \frac{ \phi'(u)^2}{u \phi''(u) - \phi'(u)} \phi'(u)^2 |_{u=0} \;.
\end{align}
This gives the formula for the $q$-th cumulant (for $q \geq 3$) coming from the microscopic
regime, to leading order in $\sigma$. We can check that for $q=3$ it gives
\be 
\langle {\cal L}_N^3 \rangle_c \simeq \sqrt{N} \frac{R}{2 \xi} \sigma^2 \phi'(0)^2 \;,
\ee 
which coincides with the leading term in \eqref{thirdcum}. 


We now compare with the formula \eqref{cumul_macro} from smooth macroscopic linear statistics.
Let us insert $f(r)= \phi(u)$ in \eqref{cumul_macro}, where $u= (r-R) \sqrt{N}/\xi$, i.e., $r = R + \frac{u \,\xi}{\sqrt{N}}$. The denominator of $A(r)$ in \eqref{def_A} becomes 
\be 
f''\left(r\right) \left(r^{-1} -
   U'\left(r\right)\right)+f'\left(r\right)
   \left( U''\left(r\right)+\frac{1}{r^2} \right) 
   \simeq \frac{\sqrt{N}}{\xi} (U''(R) + \frac{1}{R^2} ) (\phi'(u) - u \phi''(u)) \;,
\ee 
where we note that the leading terms have canceled. Hence we find, to leading order in $N$, using $U'(R) R=1$
and \eqref{xidef}
\be 
A(r) \simeq \frac{\sqrt{N}}{\xi} \frac1{U''(R) + \frac{U'(R)}{R}  } \frac{\phi'(u)^2}{ \phi'(u) - u \phi''(u) } 
= 2 \sqrt{N} \sigma^2 \xi \frac{\phi'(u)^2}{ \phi'(u) - u \phi''(u) } \;.
\ee 
Using $\partial_r = \frac{\sqrt{N}}{\xi} \partial_u $ we find 
\be 
A(r) \partial_r = 2 N \sigma^2 \frac{\phi'(u)^2}{ \phi'(u) - u \phi''(u) }  \partial_u  \;. 
\ee 
Inserting into \eqref{cumul_macro} and combining all factors we obtain exactly the formula 
\eqref{CumqMicro} (note that the factor $r$ in \eqref{cumul_macro} can be replaced
by $R$ to this order in $N$).


\section{Constrained density in the limit $\mu \to -\infty$} \label{App:const_density}

In this Appendix, we provide the analysis of the constrained density, 
specifying to the Ginibre case $U(r) = r^2/2$ and restricting to the bulk $\hat r < R$. We will focus in particular in the limit $\mu \to -\infty$. 
We start from the formula \eqref{def_rhomu} for the constrained density, which we recall here
\begin{align} \label{def_rhomu_app}
\overline{\rho}_{N}(r)\Big \vert_{U \to U + \frac{\mu}{2 N} \phi } &\simeq \rho_\mu \left(\sqrt{N}\frac{r-\hat r}{\xi} \right) \\
\rho_\mu(u) &= \frac{\sigma^2}{\pi} e^{-\frac{u^2}{2 \sigma^2}- \mu \phi(u)} \int_{-\infty}^{+\infty} d\tilde \lambda \frac{e^{\tilde \lambda u}}{\int_{-\infty}^{+\infty} du \, e^{-\frac{u^2}{2 \sigma^2}+\tilde \lambda u - \mu \phi(u)}} \;.
\end{align}
%
%
From this density the CGF $\tilde \chi(\mu,N)$ can be computed from the formula 
\bea \label{dchitilde}
\partial_\mu \tilde \chi(\mu,N) \simeq -\sqrt{N} 2 \pi \, \hat r \xi \int_{-\infty}^\infty du \rho_\mu(u) \phi(u) \;. 
\eea
One can check that this formula is consistent with the starting formula given for $\tilde \chi(\mu,N)$ in Eq.~(\ref{chi_tilde_ratio}).

We now investigate this formula (\ref{def_rhomu}) in different asymptotic regimes, in order to complement
and gain insight into the analysis performed in the main text on $\tilde \chi(\mu,N)$ for $\mu \to - \infty$.

\vspace*{0.5cm}
\noindent{\bf The limit $\mu \to - \infty$ with $\sigma$ fixed.} To simplify the discussion, we assume here that $\phi_{\max} = \phi(0) > 0$ and that $\phi''(0)$ is finite. In this case, there are two different regimes of interest: (i) $u = O(1/\sqrt{|\mu}|)$ and (ii) $u = O(\sqrt{|\mu|})$. 

\begin{enumerate}
\item[(i)] In the first regime one finds 
\begin{align} \label{first_regime}
\rho_{\mu}(u) &\sim \tilde \rho_I(v = u \sqrt{|\mu|}) \nonumber \\
\tilde \rho_I(v) &= \frac{\sigma^2}{2 \pi} |\mu \phi''(0)| \left({\rm erf}\left( \frac{\zeta_c + |\phi''(0)| v}{\sqrt{2 |\phi''(0)|}}\right) - {\rm erf}\left( \frac{-\zeta_c + |\phi''(0)| v}{\sqrt{2 |\phi''(0)|}}\right) \right)
\end{align}
where we recall that $\zeta_c = \sqrt{2 \phi(0)/\sigma^2}$ [see Eq. (\ref{large_dev})]. One can check, by inserting this asymptotic form (\ref{first_regime})
in Eq. (\ref{dchitilde}), that one recovers the asymptotic behavior of $\tilde \chi(\mu,N)$ obtained in Eq. (\ref{chi_mu_large_bulk2}) by a quite different method. 

\item[(ii)] To analyse the second regime, we use the asymptotic behavior of the integral in the denominator of Eq. (\ref{def_rhomu}), namely
\begin{align} \label{asympt_int}
\int_{-\infty}^\infty du \, e^{-\frac{u^2}{2 \sigma^2}+\zeta \sqrt{|\mu|} u + |\mu| \phi(u)} \underset{\mu \to -\infty}{\simeq} 
\begin{cases}
&\sqrt{\frac{2 \pi}{|\mu \phi''(0)|}} e^{\frac{\zeta^2}{2 |\phi''(0)|} + |\mu| \phi(0)} \quad, \quad |\zeta| \leq \zeta_c \\
& \\
& \sqrt{2 \pi \sigma ^2}\, e^{\frac{1}{2} |\mu| \sigma^2 \zeta^2} \quad, \hspace*{1.3cm}\quad |\zeta| \geq \zeta_c
\end{cases} \;.
\end{align}
\end{enumerate}
In this regime it turns out that the integral over $\tilde \lambda$ in (\ref{def_rhomu}) is dominated by the region $\tilde \lambda \geq  \zeta_c \sqrt{|\mu|}$. Performing that integral over $\tilde \lambda$, one finds finally that in this regime $\rho_\mu(u) \simeq 1/\pi \,\theta(u_c - u)$ where $u_c = \sigma^2 \zeta_c \sqrt{|\mu|}$. More precisely, one finds that $\rho_\mu(u)$ takes the scaling form
\bea \label{secondregime}
\rho_\mu(u) \simeq \tilde \rho_{II}\left( \frac{u_c-u}{\sqrt{2 \sigma^2}} \right) \quad, \quad \tilde \rho_{II}(v) = \frac{1}{2 \pi} {\rm erfc}(v) \quad, \quad u_c = \sigma^2 \zeta_c \sqrt{|\mu|} \;.
\eea
In Fig. \ref{fig_density} we illustrate these two regimes described respectively by Eq. (\ref{first_regime}) and~(\ref{secondregime}). 

\begin{figure}[t]
\centering
\includegraphics[width = 0.7 \linewidth]{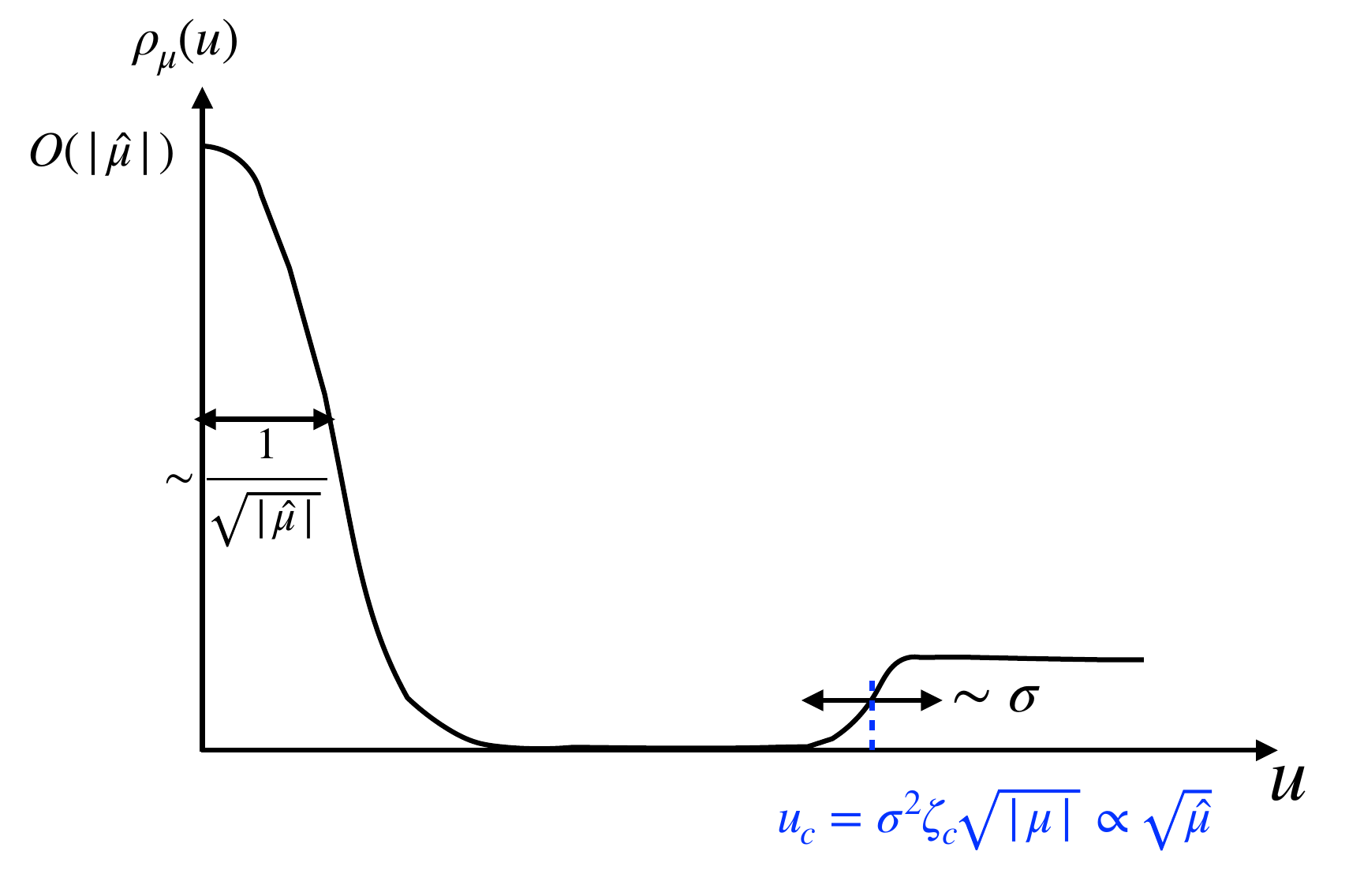}
\caption{Sketch of the constrained density profile $\rho_\mu(u)$, illustrating the two main regimes (i) and (ii) described respectively by Eq. (\ref{first_regime}) and (\ref{secondregime}).} \label{fig_density}
\end{figure}

Interestingly, one can check that the two density profiles $\tilde \rho_I(u)$ and $\tilde \rho_{II}(u)$ have a well defined limit $\mu \to - \infty$, $\sigma \to 0$ with $\hat \mu = \mu \sigma^2$ fixed. In this limit one finds (for $u>0$)
\begin{align} 
&\tilde\rho_I(u) \simeq \frac{|\hat \mu| |\phi''(0)|}{\pi} \, \theta(u_{c,1}-u) \quad, \quad u_{c,1} = \frac{1}{|\phi''(0)|} \sqrt{\frac{2 \phi(0)}{|\hat \mu|}} \label{uc1}\\
&\tilde \rho_{II}(u) \simeq \frac{1}{\pi} \theta(u-u_{c,2}) \quad, \quad \hspace*{1.3cm} u_{c,2} = \sqrt{2 \phi(0) |\hat \mu|} \;. \label{uc2}
\end{align}

\vspace*{0.5cm}
\noindent{\bf The limit $\mu \to - \infty$, $\sigma \to 0$ with $\hat \mu = \mu \sigma^2$ fixed.} We start from the formula in Eq.~(\ref{def_rhomu}) and perform the change of variable $\hat \lambda = \sigma^2 \tilde \lambda$. One gets
\begin{align} \label{def_rhomu_app}
\rho_\mu(u) = \frac{1}{\pi} e^{-\frac{1}{\sigma^2}(\frac{u^2}{2} - |\hat \mu| \phi(u))} \int_{-\infty}^\infty d \hat \lambda \frac{e^{\frac{1}{\sigma^2} \hat \lambda u}}{\int_{-\infty}^{\infty} dv \, e^{-\frac{1}{\sigma^2}(\frac{v^2}{2} - \hat \lambda v - |\hat \mu| \phi(v))}} \;.
\end{align}
In the limit $\sigma \to 0$, the integrals over $v$ and $\hat \lambda$ can be performed by a saddle point computation. The evaluation of the integral over $v$ yields
\bea \label{denominator}
\int_{-\infty}^{\infty} dv \, e^{-\frac{1}{\sigma^2}(\frac{v^2}{2} - \hat \lambda v - |\hat \mu| \phi(v))} \sim \sqrt{\frac{2 \pi \sigma^2}{1 - |\hat \mu|\phi''(v^*)}} \, e^{- \frac{1}{\sigma^2} (\frac{(v^*)^2}{2} - \hat \lambda \hat v^* - |\hat \mu|\phi(v^*))}
\eea
where $v^* \equiv v^*(\hat \lambda) = \underset{v \in \mathbb{R}}{\rm argmin}(v^2/2 - \lambda v - |\hat \mu| \phi(v))$ satisfies
\bea \label{def_vstar}
v^* - |\hat \mu| \phi'(v^*) = \hat \lambda \quad, \quad 1 - |\hat \mu| \phi''(v^*) > 0 \;. 
\eea
The study of these relations (\ref{def_vstar}) shows that the function 
$v^*(\hat \lambda)$ exhibits different behaviors depending on the sign of the function $1 - |\mu| \phi''(v)$ as a function of $v$ on the real axis. 

To see this, it is useful to consider the function $F(v; \hat \lambda)$, viewed as a function of $v$ and depending on the parameter $\hat \lambda$,   
defined as
\bea \label{def_F}
F(v; \hat \lambda) = \frac{v^2}{2} - \hat \lambda v - |\hat \mu| \phi(v) \;.
\eea
\begin{figure}[t]
\centering
\includegraphics[width =  0.9\linewidth]{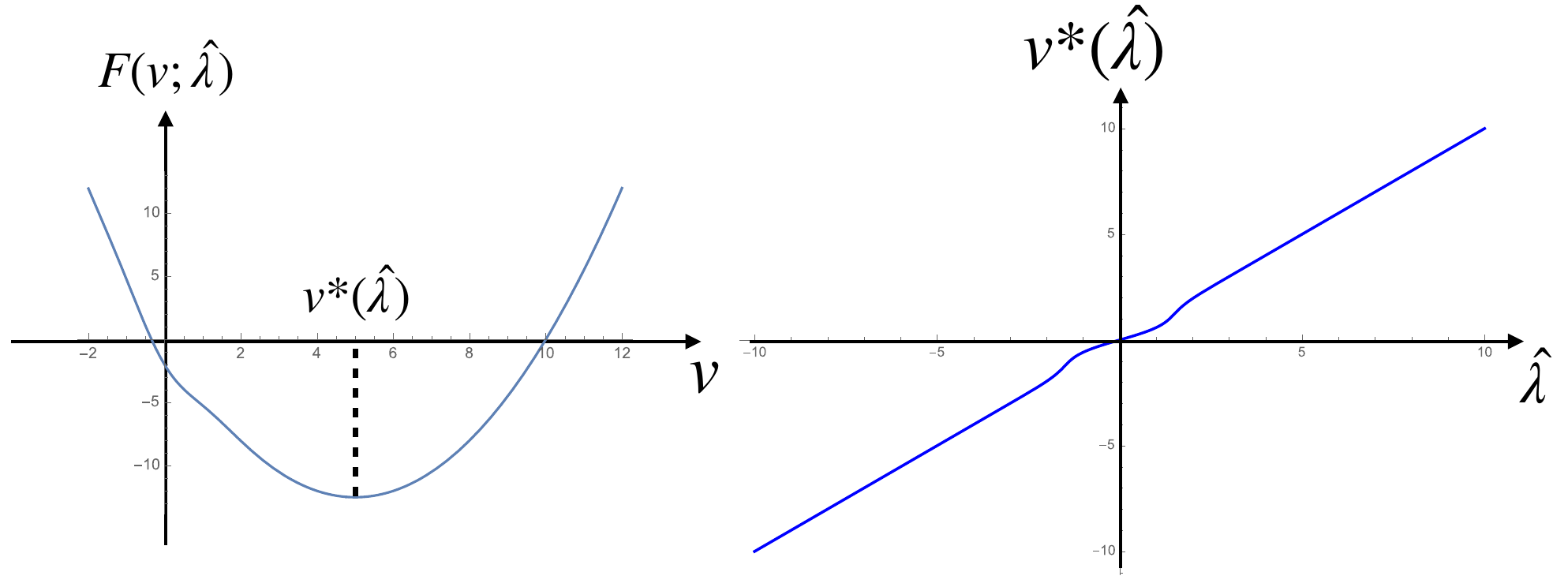}
\caption{Illustration of the case where $1 - \hat \mu \phi''(v) > 0$ for all $v \in \mathbb{R}$. {\bf Left}: Plot of $F(v;\hat \lambda)$ as a function of $v$, which for all $\hat \lambda$ exhibits a single minimum at $v^*(\hat \lambda)$. {\bf Right}: Plot of $v^*(\hat \lambda)$ as a function of $\hat \lambda$.}\label{small_mu} 
\end{figure}
We now discuss two cases:
\begin{itemize}
\item[(i)]{$1 - |\hat \mu| \phi''(v) > 0$ for all $v \in \mathbb{R}$ -- or equivalently $|\hat \mu| > |\hat \mu_c|$, for some $|\hat \mu_c|$. In this case, $F(v;\hat \lambda)$ has a unique minimum, for all values of $\hat \lambda \in \mathbb{R}$. Hence, there is a unique solution to Eq. (\ref{def_vstar}) and it is easy to show that $v^*(\hat \lambda)$ is a monotonically increasing function of $\hat \lambda$, such that $\lim_{\hat \lambda \to \pm \infty} v^*(\hat \lambda) = \pm \infty$ (see Fig. \ref{small_mu}).}
\item[(ii)]{There exists an interval $[v_1, v_2]$ where $1 - |\hat \mu| \phi''(v) < 0$ -- or equivalently $|\hat \mu| < |\hat \mu_c|$. In this case, the function $F(v;\hat \lambda)$ typically admits two local minima $v_1^* \equiv v_1^*(\lambda)$ and $v_2^* = v_2^*(\hat \lambda)$ -- both satisfying the relations in Eq. (\ref{def_vstar}), separated by a local maximum. One can then show that, for $\hat \lambda < \hat \lambda_c$, the global maximum is at $v_1^*$ while for $\lambda > \lambda_c$ it is at $v_2^*$ (see Fig. \ref{large_mu}). 
Exactly at $\hat \lambda = \hat \lambda_c$ the two local minima reach the same value at $v_1^* = u_{c,1}$ and $v_2^* = u_{c,2}$ (see Fig. \ref{large_mu}). This is the scenario of a first-order phase transition and this implies that, in this case, the function $v^*(\lambda)$ is discontinuous at $\hat \lambda = \hat \lambda_c$, with a jump from $\lim_{\hat \lambda \to \hat \lambda_c^-} v^*(\lambda) = u_{c,1}$  to $\lim_{\hat \lambda \to \hat \lambda_c^+} v^*(\lambda) = u_{c,2}$ (see Fig. \ref{vstar_vs_mu}).

To determine $u_{c,1}$, $u_{c,2}$ and $\hat \lambda_c$ we need to solve 
\begin{align}
&\frac{\partial F(v;\hat \lambda_c)}{\partial v} \Big \vert_{v=u_{c,1}} = \frac{\partial F(v;\hat \lambda_c)}{\partial v} \Big \vert_{v=u_{c,2}} = 0 \label{eq1}  \\
& F(u_{c,1}; \hat \lambda_c) = F(u_{c,2}; \hat \lambda_c)  \label{eq2} \;. 
\end{align}
Interestingly, $\hat \lambda_c$ can be eliminated from the first relation (\ref{eq1}), leading to
\begin{align} \label{lambdac}
\hat \lambda_c = u_{c,1} - |\hat \mu| \phi'(u_{c,1}) =  u_{c,2} - |\hat \mu| \phi'(u_{c,2}) \;,
\end{align}
which also provides a first relation between $u_{c,1}$ and $u_{c,2}$. Furthermore substituting this expression of $\hat \lambda_c$ in terms of $u_{c,1}$ and $u_{c,2}$ in Eq. (\ref{eq2}) one finds a second relation between $u_{c,1}$ and $u_{c,2}$, namely
\begin{equation} \label{second_rel}
- \frac{u_{c,1}^2}{2} + |\hat \mu| (u_{c,1} \phi'(u_{c,1})- \phi(u_{c,1})) = - \frac{u_{c,2}^2}{2} + |\hat \mu| (u_{c,2} \phi'(u_{c,2})- \phi(u_{c,2})) \;. 
\end{equation}
Hence these two equations (\ref{lambdac}) and (\ref{second_rel}) can be solved numerically to obtain $u_{c,1}$ and $u_{c,2}$, while $\hat \lambda_c$ can be obtained from Eq. (\ref{lambdac}). In addition, by analysing these two equations (\ref{lambdac}) and (\ref{second_rel}) in the large $|\hat \mu|$ limit, one recovers the behaviors found in Eqs. (\ref{uc1}) and (\ref{uc2}), as it should.  
}
\end{itemize}

\begin{figure}[t]
\centering
\includegraphics[width=\linewidth]{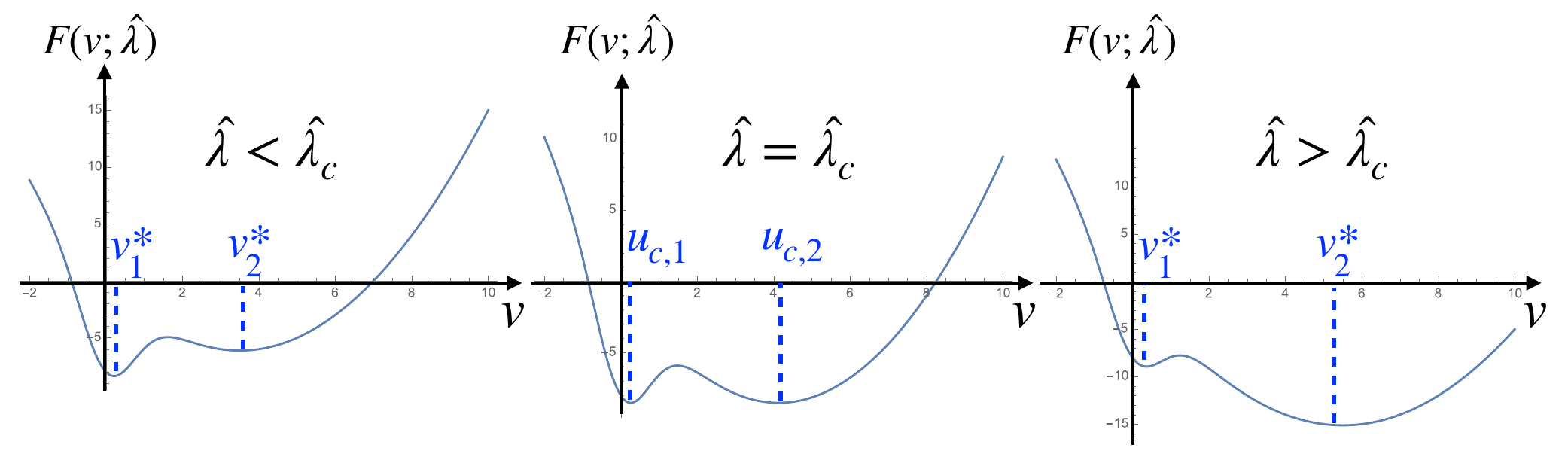}
\caption{Plot of $F(v;\hat \lambda)$ vs $v$ in the case where $1 -|\hat \mu| \phi''(v) < 0$ for $v \in [v_1, v_2]$ for different values of $\hat \lambda$, as explained in the text.} \label{large_mu}
\end{figure}
To proceed in the evaluation of $\rho_\mu(u)$ given in (\ref{def_rhomu_app}) in the limit $\sigma \to 0$ keeping $\hat \mu = \sigma^2 \mu$ fixed, we insert the asymptotic behavior of the denominator (\ref{denominator}) in (\ref{def_rhomu_app}) and evaluate the integral over $\hat \lambda$ via a saddle point. The saddle point corresponds to the maximum of $\hat \lambda u + F(v^*(\hat \lambda), \hat \lambda)$ and the stationarity condition at the saddle point $\hat \lambda = \hat \lambda^*$ reads
\begin{align} \label{stat}
v^*(\hat \lambda ^*) = u \;. 
\end{align}
From the previous analysis, we thus see that the saddle point always exists for $|\hat \mu| < |\hat \mu_c|$ (see Fig. \ref{small_mu}), while, for $|\hat \mu| > |\hat \mu_c|$ it exists only for $u<u_{c,1}$ and $u>u_{c,2}$. Performing carefully the saddle point computation finally yields the following results.

\begin{enumerate}
\item[]{For $|\hat \mu| < |\hat \mu_c|$, the density is a smooth function given by
\begin{align}
\rho_\mu(u) \simeq \frac{1}{\pi} \left(1 - |\hat \mu| \phi''(u) \right) \quad, \quad \forall \; u \geq  0 \;. 
\end{align}
}
\item[]{For $|\hat \mu| > |\hat \mu_c|$, the density $\rho_{\mu}(u)$ exhibits a hole for $u_{c,1}< u < u_{c,2}$, i.e.,  
\begin{align} \label{hole}
\rho_{\mu}(u) \simeq
\begin{cases}
&\frac{1}{\pi} \left(1 - |\hat \mu| \phi''(u) \right) \quad, \quad 0<u<u_{c,1} \\
& 0 \quad, \quad \hspace*{2.7cm} u_{c,1}<u<u_{c,2} \\
& \frac{1}{\pi} \left(1 - |\hat \mu| \phi''(u) \right) \quad, \quad u_{c,2}<u
\end{cases}
\end{align}
where $u_{c,1}$ and $u_{c,2}$ are given by solving Eqs. (\ref{lambdac}) and (\ref{second_rel}). In the limit $|\hat \mu \vert \to \infty$ one can check that (\ref{hole}) reproduces the result in Eqs. (\ref{uc1}) and (\ref{uc2}). 
}
\end{enumerate}

\noindent {\bf Remark 9}. One can also estimate the constrained density from the saddle point
approach used to obtain \eqref{chi_small_sigma} in the limit $\sigma \to 0$ with 
$\hat \mu=\sigma^2 \mu$ fixed. To leading order it gives 
\be
\rho_\mu(u) \sim \exp\left(  - \frac{1}{\sigma^2} \left( \min_{\hat \lambda \in \mathbb{R}}( \frac{u^2}{2} - \hat \lambda u + \hat \mu \phi(u) 
- \min_{u \in \mathbb{R}} \left[ \frac{u^2}{2} - \hat \lambda u + \hat \mu \phi(u) \right]  \right) \right) 
\ee 
which also predicts the regions where the density vanishes as $\sigma \to 0$.
To obtain the prefactor and recover \eqref{hole} one would need to take
into account the fluctuations around the saddle point.

\begin{figure}[t]
\centering
\includegraphics[width = 0.6\linewidth]{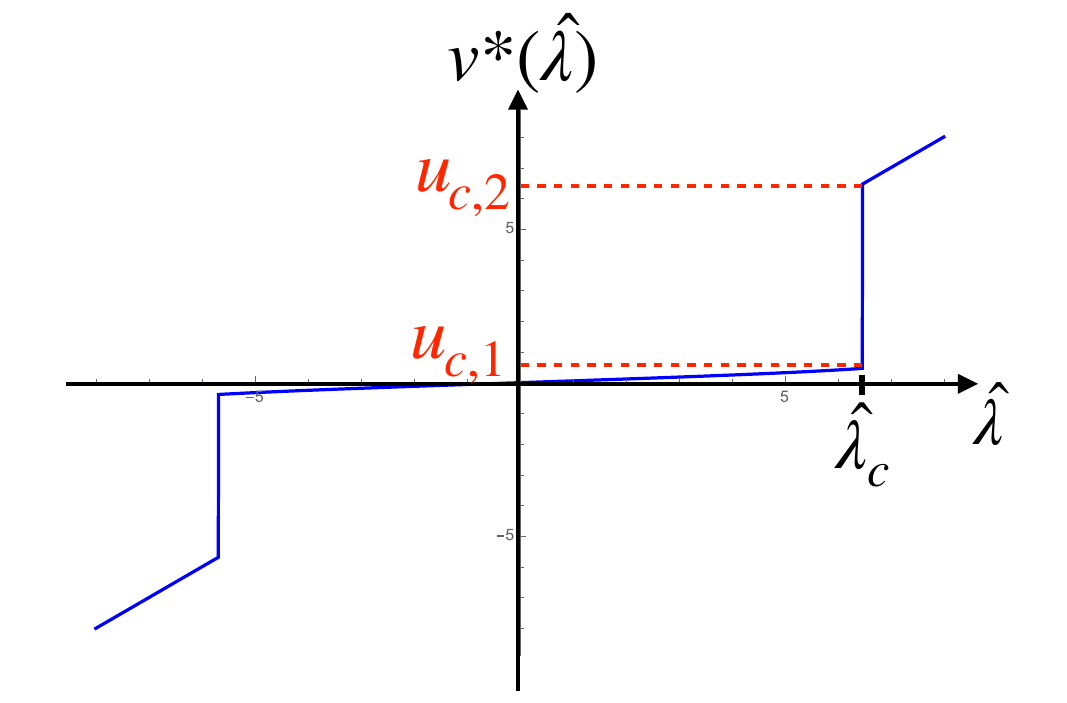}
\caption{Plot of $v^*(\hat \lambda)$ vs $\hat \lambda$ in the case $1 -|\hat \mu| \phi''(v) < 0$ for $v \in [v_1, v_2]$. It exhibits a jump at $\hat \lambda = \hat \lambda_c$ from $u_{c,1}$ to $u_{c,2}$ -- see Fig. \ref{large_mu}.} \label{vstar_vs_mu}
\end{figure}
\section{Coulomb gas approach} \label{sec:CG}


In this Appendix we solve the minimization problem defined in Eqs. \eqref{energy}-\eqref{def_I}.
We assume that the maximum of $\phi(u)$ is located at $u=0$, hence $\phi_{\rm max}=\phi(0)$. Here for simplicity, we assume here that $\phi(u) \geq 0$ for all $u \in \mathbb{R}$ (and in particular $\phi(0)>0$) and we restrict our analysis to the case of the Ginibre ensemble, $U(r)=r^2/2$ (althgouh in principle it can be extended to more general potentials). Let us define 
\be
\tilde \Lambda = \frac{\Lambda}{\phi(0)}  \quad , \quad 0 \leq \tilde \Lambda < 1 \;.
\ee 
Let us start with the normalisation condition \eqref{norm_CG} for the density $\rho({\bf r})$. From the form (\ref{rho_ansatz}) one obtains the following relation
\be
\tilde \Lambda - (r_2^2-r_1^2) = 1-R^2
\ee
We have checked that allowing for a general $R$ and minimizing the energy w.r.t. $R$ we obtain as possible solutions either (i) $R=1$ or (ii) 
$r_2=R$. The second solution simply means that the annulus is reduced to zero, hence the support is a
a disk of radius $r_1$ with $r_1^2= 1- \tilde \Lambda$ by normalization (in this case both $r_2$ and $R$ thus disappear from the optimization problem). Hence we can safely set $R=1$ in the following. 

After a tedious calculation one finds that the energy reads 
\begin{align} \label{energy0}
{\cal E}[\rho] = &\; \hat r^2  \frac{\tilde \Lambda}{2} - \frac{1}{4} (r_2^4-r_1^4 -1)  \\
&- \frac{1}{2} \tilde \Lambda^2  \log \hat r  - \tilde \Lambda \left({r_1^2} \ln \hat r + \frac{1}{2}\left( -1+r_2^2-2 r_2^2 \ln r_2\right) \right) \nonumber  \\
& + \frac{1}{8} \left(r_1^4+4
   r_1^2+3 r_2^4-4
   \left(r_1^2+1\right)
   r_2^2-4 r_1^4 \log
   \left(r_1\right)+\left(8
   r_1^2 r_2^2-4 r_2^4\right)
   \log
   \left(r_2\right)+1\right) \;. \nonumber 
\end{align} 
These computations can be easily performed using the identity (for $a,b>0$)
\bea \label{identity}
\int_0^{2 \pi} \ln (a^2 + b^2 + 2 ab \cos{\theta}) d\theta = 4 \pi \ln\left(\max(a,b)\right) \;.
\eea
Let us enforce the normalization constraint and substitute $r_2^2 = r_1^2 + \tilde \Lambda$
in \eqref{energy0}. The energy simplifies into 
\begin{align} \label{E_Simplif}
&{\cal E}[\rho] -  {\cal E}[\rho_{\rm eq}] = {\cal E}[\rho] - \frac{3}{8}  \\
= 
&\frac{1}{8} \Big(2
   \left(\tilde \Lambda
   +r_1^2\right){}^2 \log
   \left(\tilde \Lambda
   +r_1^2\right)+\tilde \Lambda 
   \left(-3 \tilde \Lambda -6 r_1^2+4
   {\hat r}^2\right)-2
   \tilde \Lambda  \left(\tilde \Lambda +2
   r_1^2\right) \log
   \left({\hat r}^2\right) \nonumber \\
   &-2
   r_1^4 \log
   \left(r_1^2\right)\Big) \;,  \nonumber 
\end{align} 
where we have used ${\cal E}[\rho_{\rm eq}] = 3/8$. One can check that this function is a convex function of $r_1$. To minimize the energy, one takes a derivative w.r.t. $r_1$ and obtains an equation for $r_1$ which takes the form
\be  \label{eq_f}
r_1^2 = \hat r^2 f(z)  \quad , \quad z = \frac{\tilde \Lambda}{\hat r^2} \;, \qquad {\rm with} \quad
(z + f(z) ) \log(z + f(z)) - f(z) \log f(z) = z  \;.   \ee 
The function $f(z)$ can be obtained, e.g. in an expansion at small $z$, which reads
\be \label{small_z} 
f(z) = 1 - \frac{z}{2} + \frac{z^2}{24}  + \frac1{1920 } z^4 + \dots 
\ee 
Note that $0 \leq  f(z)<1$, consistent with $r_1<\hat r$. 
One can also check that $f(z)+z > 1$, consistent with $r_2>\hat r$. 
Substuting the optimal value of $r_1$ into the energy and using the equation satisfied by $f(z)$, namely Eq. (\ref{eq_f}),
one obtains the optimal energy as 
\bea 
&& {\cal E}[\rho_\Lambda] - \frac{3}{8} = \hat{r}^4 {\sf E}\left(\frac{\Lambda}{\phi(0) \hat r^2}\right) \nonumber \\
{\sf E}(z) &=& \frac{z}{8} \left( 2 f(z) \log f(z) - 4 f(z) + (4-z)  \right) \label{defE_1}\\
&=& \frac{z}{8} \left( 2(z+f(z))\log(z+f(z)) - 4 f(z)+ 4 - 3z\right) \;. \label{defE_2}
\eea 
This result is only valid when the optimal values for $r_1$ and $r_2$ satisfy $0<r_1<r_2<1$, 
which is the white region in Fig. \ref{Fig_PhDiag}. 

In the green region in Fig. \ref{Fig_PhDiag}, the optimal solution is instead given by $r_1=0$. The frontier of the domain 
is given by $\tilde \Lambda=\tilde \Lambda_1$, together with the condition $\tilde \Lambda <1$, with
\be \label{tildelambda1}
\tilde \Lambda_1  =  e \hat r^2  \quad , \quad z=z_1 = e \quad , \quad \hat r^2 < 1/e \;.
\ee
This is obtained by inserting the condition $f(z_1)=0$ into \eqref{eq_f}. 
One can check that the result \eqref{defE_1}-\eqref{defE_2} is then valid only for $\tilde \Lambda < \tilde \Lambda_1 $, i.e.
$z<z_1$. In the green region, substituting $r_1=0$ into \eqref{E_Simplif} one 
obtains the optimal energy as
\bea \label{E1}
{\cal E}[\rho_\Lambda] - \frac{3}{8}  = \hat r^4 {\sf E}_1(z) \quad, \quad {\sf E_1}(z) = \frac{z}{8} \left( 2 z \log z + 4 - 3z\right) \quad, \quad z > e \
\eea


In the blue region in Fig. \ref{Fig_PhDiag}, the optimal solution is given by $r_2=1$, i.e. $r_1^2=1 -\tilde \Lambda$. The frontier of the domain is obtained by substituting $r_1^2=1 -\tilde \Lambda$
in \eqref{eq_f}, together with $\tilde \Lambda <1$. It is thus
given by $\tilde \Lambda= \tilde \Lambda_2$, where $\tilde \Lambda_2$ is solution of the equation
\be \label{eq_lambda2}
- \left(\frac{1}{\tilde \Lambda_2}- 1\right) \log(1- \tilde \Lambda_2) = 1 + \log \hat r^2  \quad , \quad \hat r^2 > 1/e \;.
\ee 
The solution can be obtained explicitly in terms of the Lambert $W$-function 
\bea  \label{lambda2}
\frac{1}{1 - \tilde \Lambda_2} = - \frac{1}{A} W_{k}(- A e^{-A}) \quad, \quad  A= 1 + \log \hat r^2 = \log \tilde \Lambda_1
\eea 
where one should take the branch parametrized by $k=-1$ since here $A \leq 1$. In this case, one finds that the optimal energy is given by
\begin{align} \label{E2}
&{\cal E}[\rho_\Lambda] - \frac{3}{8} \\
&= 
\frac{1}{8}\left(\tilde \Lambda(3 \tilde \Lambda + 4 \hat r^2 - 6) + (2 \tilde \Lambda^2-4 \tilde \Lambda) \ln \hat r^2 - 2 (1-\tilde \Lambda)^2 \log(1-\tilde \Lambda) \right) \quad, \quad \tilde \Lambda_2 < \tilde \Lambda < 1 \nonumber 
\end{align}

In summary we obtain the rate function $\Psi_{\rm CG}(\Lambda)$ defined in
the text in \eqref{Psi_CG} as follows:

For $\hat r^2 < 1/e$, the rate function $\Psi_{\rm CG}(\Lambda)$ exhibits a transition at $\Lambda = \phi(0)\tilde \Lambda_1 = e \hat r^2 \phi(0)$, and reads 
\bea 
\Psi_{\rm CG}(\Lambda) = 
\begin{cases} \label{psi_CG_inf}
&\beta \hat r^4 {\sf E}\left(\frac{\Lambda}{\phi(0)\hat r^2} \right)\quad, \quad 0<\Lambda < \phi(0) e \hat r^2 \\
&\beta \hat r^4 {\sf E}_1\left(\frac{\Lambda}{\phi(0)\hat r^2} \right)\quad, \quad \phi(0) e \hat r^2 < \Lambda < \phi(0)
\end{cases}
\eea
where ${\sf E}(z)$ is given in Eq. (\ref{defE_1}), or equivalently in (\ref{defE_2}), where $f(z)$
is the solution of \eqref{eq_f},
while ${\sf E}_1(z)$ is given in (\ref{E1}).

For $\hat r^2 > 1/e$, the rate function $\Psi_{\rm CG}(\Lambda)$ exhibits a transition at $\Lambda = \phi(0)\tilde \Lambda_2$ where $\tilde \Lambda_2$ is defined in (\ref{lambda2}). One has in this case
\begin{align} \label{transition_E2}
\Psi_{\rm CG}(\Lambda) = 
\begin{cases}
&\beta \hat r^4 {\sf E}\left(\frac{\Lambda}{\phi(0)\hat r^2} \right)\quad, \quad 0<\Lambda < \phi(0) \tilde \Lambda_2 \\
&\frac{\beta}{8}\left(\tilde \Lambda(3 \tilde \Lambda + 4 \hat r^2 - 6) + (2 \tilde \Lambda^2-4 \tilde \Lambda) \ln \hat r^2 - 2 (1-\tilde \Lambda)^2 \log(1-\tilde \Lambda) \right) \\
& \hspace*{6.5cm}\quad {\rm for}\quad \phi(0) \tilde \Lambda_2 < \Lambda < \phi(0) \;,
\end{cases}
\end{align}
where we recall that $\tilde \Lambda = \Lambda/\phi(0)$. 

Let us now analyze the asymptotic behaviors of the rate function, and determine the order of the various transitions. 

\vspace*{0.3cm}
\noindent{\bf The limit $\Lambda \to 0$}. The limit $\Lambda \to 0$ can be obtained by analysing the expression of ${\sf E}(z)$ given in Eq. (\ref{defE_1}) or (\ref{defE_2}) in the limit $z \to 0$, using the small $z$ behavior of $f(z)$ given in Eq. (\ref{small_z}). One finds 
\bea 
\Psi_{\rm CG}(\Lambda) \simeq \frac{1}{24} \frac{\Lambda^3}{\phi(0)^3 \hat r^2} \left( 1 + 
\frac{\Lambda^2}{240 \phi(0)^2 \hat r^4} + \dots \right)  \quad, \quad \Lambda \to 0 \;,
\eea 
which coincides with the result from microscopic scales obtained in Eqs. (\ref{cubic}) and~(\ref{Apm}).

\noindent{\bf The behavior close to $\Lambda = \Lambda_1 = \phi(0) e \hat r^2$}. To analyse the transition between the two regimes in Eq. (\ref{psi_CG_inf}), we first analyse the behavior of ${\sf E}(z)$ as $z \to e$, with $z<e$. Using the expansion of $f(z)$ for $z \to e$ (with $f(e) = 0$) as
\bea \label{asympt_e}
f(z) \simeq \frac{(e-z)}{|\log(e-z)|} - (e-z) \frac{\log(|\log(e-z)|)}{\log^2(e-z)} + \cdots \quad, \quad z \to e \;. 
\eea
one finds 
\be \label{asympt_E_e}
{\sf E}(z) \sim \frac{1}{8}(4e-e^2) + \frac{z-e}{2} + \frac{1}{4}(z-e)^2  - \frac{1}{4}\frac{(e-z)^2}{|\ln(e-z)|} + \cdots \qquad, \qquad z \to e \; {\rm with} \;  z<e \;.  
\ee
On the other hand, for $z>e$, one finds that ${\sf E}_1(z)$ behaves, for $z \to e$, as 
\be \label{asympt_E1_e}
{\sf E}_1(z) \sim \frac{1}{8}(4e-e^2) + \frac{z-e}{2} + \frac{1}{4}(z-e)^2 + \frac{e^{-1}}{12}(z-e)^3+ O((z-e)^4) \quad, \quad z \to e \; {\rm with} \;  z>e \;.  
\ee
Hence, by comparing these expansions (\ref{asympt_E_e}) and (\ref{asympt_E1_e}), we see that the rate function $\Psi_{\rm CG}(\Lambda)$ in (\ref{psi_CG_inf}) as well as its first and second derivatives are continuous in $\Lambda_1 = \tilde \Lambda_1 \phi(0)=\phi(0) e \hat r^2$. Its third derivative is instead singular: it is infinite (and positive) on the left hand side while it is finite on the right hand side of $\Lambda_1 = \tilde \Lambda_1 \phi(0)=\phi(0) e \hat r^2$. 

\begin{figure}[t]
\centering
\includegraphics[width = 0.6 \linewidth]{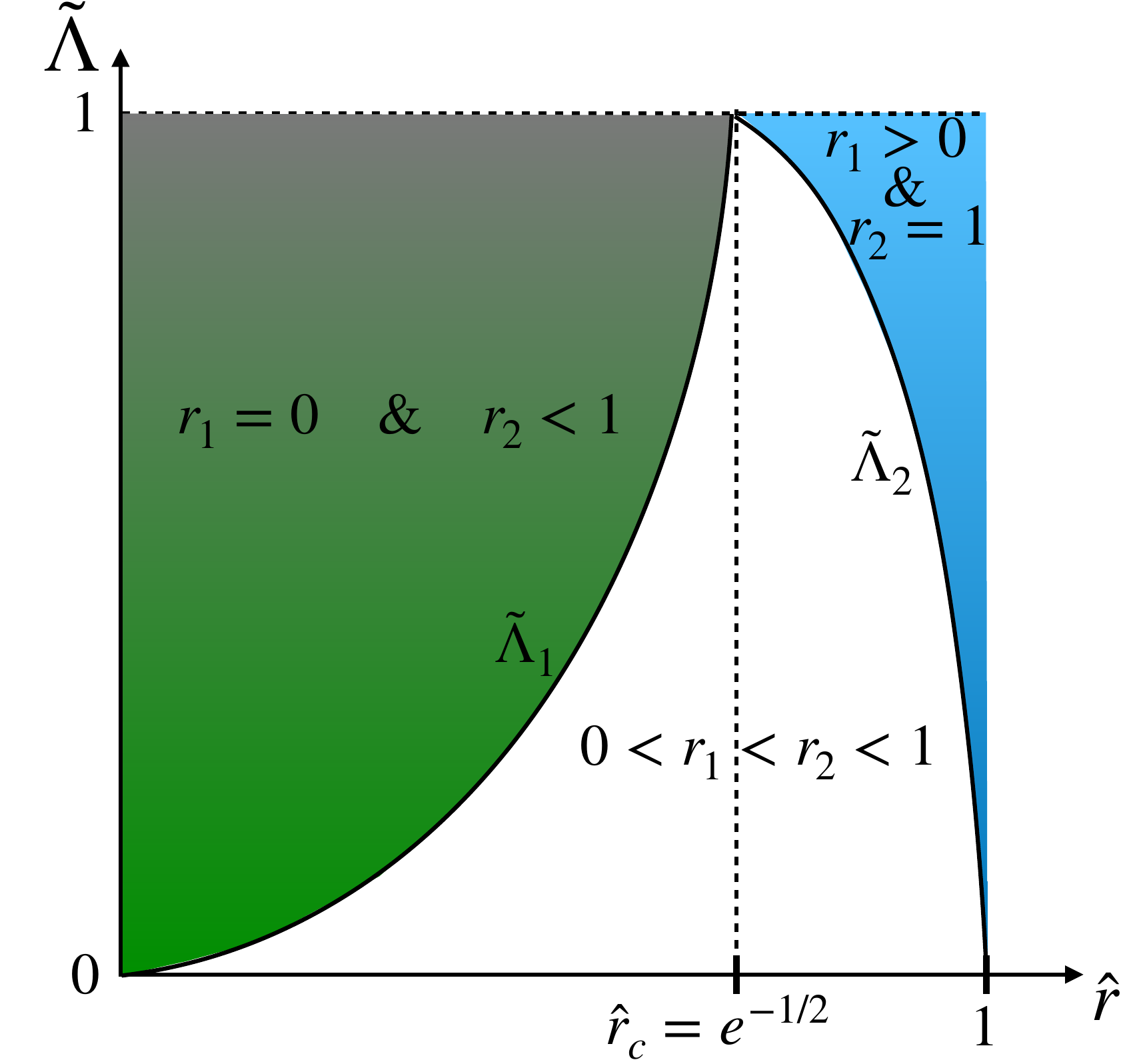}
\caption{Phase diagram in the plane $(\hat r, \tilde \Lambda = \Lambda/\phi(0))$ associated to the behavior of the rate function $\Psi_{\rm CG}(\Lambda)$ in the Coulomb gas regime, as given in Eqs (\ref{psi_CG_inf}) and (\ref{transition_E2}). The radii $r_1$ and $r_2$ characterize the conditioned density $\rho_\Lambda(r)$ defined in Eqs. (\ref{rho_ansatz})-(\ref{def_I}) and plotted in Fig. \ref{Fig_hole}. The frontiers $\tilde \Lambda_1$ and $\tilde \Lambda_2$ can be explicitly computed and they are given in Eqs. (\ref{tildelambda1}) and (\ref{lambda2}) respectively.} \label{Fig_PhDiag}
\end{figure}
\noindent{\bf The behavior close to $\Lambda = \Lambda_2 = \phi(0) \tilde \Lambda_2$}. We now analyse the behavior between the two regimes described in Eq. (\ref{transition_E2}). One can check that $\Psi_{\rm CG}(\Lambda)$ is continuous at $\Lambda = \Lambda_2 = \tilde \Lambda_2 \phi(0)$ with the value
\bea
\Psi_{\rm CG}(\Lambda_2) = \beta \frac{\tilde \Lambda_2}{8}\left(\tilde \Lambda_2 + 4(\hat r^2-1) - 2 \log \hat r^2 \right) \;.
\eea
Its first derivative is also continuous, with the result 
\bea
\Psi_{\rm CG}'(\Lambda_2) = \frac{\beta}{2}(\hat r^2 - \log \hat r^2-1) \;,
\eea
where we have used
\bea
f'\left( \frac{\tilde \Lambda_2}{\hat r^2}\right) = - \frac{\log \hat r^2}{\log(1-\tilde \Lambda_2)} = \frac{1-\tilde \Lambda_2}{\tilde \Lambda_2} \frac{\log \hat r^2}{1+\log \hat r^2} \;. 
\eea
Note that in the second equality we have used the relation (\ref{eq_lambda2}). However, one finds that the second derivative is discontinuous. After some tedious computations (done with the help of Mathematica) one finds indeed
\bea \label{psi_seconde}
\Psi''_{\rm CG}(\Lambda_2^-)=-\beta\frac{\log \hat r^2 \left(\tilde \Lambda_2+\log \hat r^2\right)}{2 \tilde \Lambda_2
   \left(\log \hat r^2+1\right)} \quad, \quad \Psi''_{\rm CG}(\Lambda_2^+) = \frac{\beta}{2} \frac{\tilde \Lambda_2 + \log \hat r^2}{1-\tilde \Lambda_2} \;,
\eea
which shows that the second derivative is discontinuous across $\Lambda_2 = \phi(0) \tilde \Lambda_2$. 

\noindent{\bf The behavior when $\Lambda \to \phi(0)$}. In this case, using the second line of Eqs. (\ref{psi_CG_inf}) and (\ref{transition_E2}) one finds at leading order 
\bea \label{leading} 
\Psi_{\rm CG}(\Lambda) \simeq \frac{\beta}{8}\left(4 \hat r^2 - 2 \ln \hat r^2 -3 \right)  \quad , \quad \Lambda \to \phi(0) \;,
\eea
which is valid for all $\hat r \in [0,1]$. One can check 
that $\tilde \Lambda=1$ (which is the maximal possible value for the linear statistics)
corresponds to an optimal density exactly
equal to the delta function at $r=\hat r$ (with no smooth charge
background), and indeed \eqref{energy0} corresponds to the energy of
such a configuration. Interestingly this energy exhibits
a minimum as a function of $\hat r$ at $\hat r = \sqrt{2}/{2}$. We note also that the coefficient 
of the correction of order $O((\phi(0)- \Lambda))$ in Eq. (\ref{leading}) has a different expression for $\hat r^2<1/e$ and $r^2>1/e$.

\end{document}